\documentclass[aps, prd, 10pt,twocolumn,superscriptaddress,nofootinbib]{revtex4-2}
\usepackage{mathrsfs, amssymb, amsmath, mathtools}  
\usepackage{cancel, comment}
\usepackage{footmisc}
\usepackage{latexsym}
\usepackage{natbib}
\usepackage{url}
\usepackage{dcolumn}
\usepackage{multirow}
\usepackage{color}
\usepackage{soul}
\usepackage[normalem]{ulem}
\usepackage{amsfonts,amssymb,amsmath, txfonts}
\usepackage{graphicx,epsfig}
\usepackage{psfrag}
\usepackage{hyperref}
\usepackage{mathtools}
\usepackage{enumitem}
\usepackage{float}
\usepackage[dvipsnames]{xcolor}
\usepackage{xcolor}
\hypersetup{linktoc=all,
    colorlinks=true, linkcolor={brightpink},
    citecolor={blue}, urlcolor={blue}
}
\definecolor{rosy}{RGB}{230,235,252}
\definecolor{myframetitle}{RGB}{90,89,170}
\definecolor{myblocktitle}{RGB}{140,185,249}
\definecolor{mytitle}{RGB}{10,80,26}

\definecolor{darkgreen}{RGB}{27,130,45}
\definecolor{darkblue}{rgb}{0,0,0.3}
\definecolor{darkred}{rgb}{0.7,0,0}

\definecolor{light gray}{RGB}{220,220,220}
\definecolor{dark purple}{RGB}{108,0,217}
\definecolor{pink}{RGB}{190,20,100}
\definecolor{orang}{RGB}{193,63,0}
\definecolor{green}{RGB}{11,98,17}
\definecolor{darkpink}{RGB}{153,0,76}
\definecolor{bluegreen}{RGB}{0,102,102}
\definecolor{greenlagan}{RGB}{0,102,0}
\definecolor{redgreen}{RGB}{102,102,0}
\definecolor{Redgreen}{RGB}{153,76,0}
\definecolor{vividviolet}{rgb}{0.62, 0.0, 1.0}
\definecolor{amaranth}{rgb}{0.9, 0.17, 0.31}
\definecolor{palatinateblue}{rgb}{0.15, 0.23, 0.89}
\definecolor{brightpink}{rgb}{1.0, 0.0, 0.5}
\definecolor{cornflowerblue}{rgb}{0.39, 0.58, 0.93}
\definecolor{deepcarminepink}{rgb}{0.94, 0.19, 0.22}
\definecolor{radicalred}{rgb}{1.0, 0.21, 0.37}
\def\H0{{\text{H}\hspace*{-2.05mm}\text{H} 0\hspace*{-1.35mm}0\ }}

\def\be{\begin{equation}}
\def\ee{\end{equation}}
\def\beq{\begin{equation}}
\def\eeq{\end{equation}}
\def\bea{\begin{eqnarray}}
\def\eea{\end{eqnarray}}
\newcommand{\dd}{\textrm{d}}

\begin{document}

\title{On frequentist confidence intervals in a non-Gaussian regime}

\author{Shubham Barua}
\affiliation{Department of Physics, IIT Hyderabad Kandi, Telangana 502284, India}
\author{Shantanu Desai}
\affiliation{Department of Physics, IIT Hyderabad Kandi, Telangana 502284, India}
\author{Mauricio Lopez-Hernandez}
\affiliation{Departamento de F\'isica, Centro de Investigaci\'on y de Estudios Avanzados del I.P.N.
Apartado Postal 14-740, 07000, Ciudad de M\'exico, M\'exico}
\affiliation{Atlantic Technological University, Ash Lane, Sligo F91 YW50, Ireland}
\author{Eoin \'O Colg\'ain}
\affiliation{Atlantic Technological University, Ash Lane, Sligo F91 YW50, Ireland}

\begin{abstract}
We study frequentist confidence intervals based on graphical profile likelihoods (Wilks' theorem, likelihood integration), and the Feldman-Cousins (FC) prescription, a generalisation of the Neyman belt construction, in a setting with non-Gaussian Markov chain Monte Carlo (MCMC) posteriors. Our simplified setting allows us to recycle the MCMC chain as an input in all methods, including mock simulations underlying the FC approach. We find all methods agree to within $10 \%$ in the close to Gaussian regime, but extending methods beyond their regime of validity leads to greater discrepancies. As a key consistency check, we recover a shift in cosmological parameters between low and high redshift cosmic chronometer data with the FC method, but only when one fits all parameters back to the mocks. We observe that fixing parameters, a common approach in the literature, risks underestimating confidence intervals.  
\end{abstract}

\maketitle

\section{Introduction}
If not due to systematics, $\Lambda$CDM tensions, most notably $H_0$ and $S_8$ tensions \cite{DiValentino:2021izs, Perivolaropoulos:2021jda, Abdalla:2022yfr, CosmoVerse:2025txj},  motivate the study of extended models beyond the current standard model. However, as one expands the $\Lambda$CDM parameter space, there is a danger that degeneracies between parameters may bias traditional Bayesian inference. This has led to a resurgence in interest in complementary frequentist methods, more precisely profile likelihoods (see \cite{Trotta:2017wnx, Herold:2024enb,PDG} for reviews), as a means to check Bayesian credible intervals, most notably Planck CMB analysis \cite{Planck:2013nga}, as well as to diagnose volume or projection effects in more elaborate models \cite{Gomez-Valent:2022hkb}.\footnote{See \cite{Hadzhiyska:2023wae, Gsponer:2023wpm, Kerscher:2024doc} for studies of these effects and their mitigation in the Bayesian framework.} Most of the recent interest in this direction has been fueled by a family of Early Dark Energy models \cite{Poulin:2018cxd, Agrawal:2019lmo, Lin:2019qug, Niedermann:2019olb, Ye:2020btb} and their potential to alleviate $H_0$ tension \cite{Herold:2021ksg, Herold:2022iib, Cruz:2023cxy, Goldstein:2023gnw, McDonough:2023qcu, Efstathiou:2023fbn, Poulin:2025nfb}. Given the complementarity with Bayesian methods, profile likelihoods have wide application in cosmology \cite{Hamann:2011,Campeti:2022vom, Holm:2022kkd, Giare:2023qqn, Moretti:2023drg, Planck:2013nga, Noriega:2024lzo, Desai:2024hlx, Sah:2024csa, Herold:2025hkb, Barua:2025zqo, Montandon:2025xpd, Poudou:2025qcx,  Ramakrishnan:2025biz, Chebat:2025kes}. 

Within the $\Lambda$CDM tensions context, it is important how one defines errors, since underestimated errors risk inflated tensions, and conversely overestimated errors may lead to a false sense of security. For this reason, it is interesting to compare Bayesian credible intervals with frequentist confidence intervals. The first key difference is that Bayesian analysis allows for priors that may dictate model parameter space when data is poorly constrained. Secondly, in Bayesian analysis one marginalises over nuisance parameters, and degeneracies between parameters can lead to projection effects. Obviously, given the different assumptions being made, care is required with any comparison, but in the asymptotic large sample limit, characterised by Gaussian distributions, agreement is guaranteed. Interestingly, one can find examples in the literature where Bayesian credible intervals are more conservative  than frequentist confidence intervals \cite{Colgain:2023bge, Naredo-Tuero:2024sgf} and the converse is also true \cite{Holm:2023laa, Galloni:2024lre}. This difference may in part be traced to the assumptions one imposes on nuisance parameters \cite{Holm:2023laa}. 

Even when one embraces profile likelihoods, one is confronted with a number of options of varying computational expense. The simplest proposal is to work in the large sample (Gaussian) limit and employ Wilks' theorem \cite{Wilks} to study differences in fit through $ \Delta \chi^2$. The advantage of this approach is that it only requires a knowledge of the profile likelihood in the vicinity of the maximum likelihood estimator (MLE). In the same limit, one can study the profile likelihood more globally \cite{Gomez-Valent:2022hkb}. The first complication comes with any boundary preventing a parameter from entering an unphysical regime. Here, one can piggy-back on results in the Feldman-Cousins (FC) paper for a Gaussian with a boundary \cite{Feldman:1997qc} (for example \cite{Naredo-Tuero:2024sgf}). In the terminology of \cite{Herold:2024enb}, these are \textit{graphical profile likelihoods} that are only fully valid in the large sample limit. See \cite{Herold:2024enb, Nygaard:2023cus, Holm:2023uwa, Karwal:2024qpt} for tools to determine graphical profile likelihoods and \cite{Herold:2024enb} for key tests of the large sample limit. 

More generically, one employs the Neyman confidence belt method \cite{Neyman:1937uhy}, which requires an evaluation of the likelihood for many mock realisations of the data, thereby greatly increasing the computational expense over graphical profile likelihoods that only require a single realisation of observed data. FC performs this procedure for one measured quantity with one unknown parameter \cite{Feldman:1997qc}, with the added complexity that one constructs the acceptance intervals in the final Neyman confidence belt from a ranking of likelihood ratios. FC also generalise to two parameters with one measured quantity in the context of neutrino oscillation physics \cite{Feldman:1997qc}. Throughout, we refer to this as the FC construction, which is the most general method that we consider in the paper.

Bearing in mind that Wilks' theorem, etc, are approximations that break down for smaller samples, there may still be value in a $\Lambda$CDM tensions context \cite{DiValentino:2021izs, Perivolaropoulos:2021jda, Abdalla:2022yfr, CosmoVerse:2025txj} provided they turn out to be more conservative and bound the confidence intervals above. Indeed, for Gaussian graphical profile likelihoods, one can prove that any correction due to a boundary contracts confidence intervals (see the appendix in \cite{Colgain:2024clf}). Thus, in this work we pay special attention to whether computationally inexpensive profile likelihood methods can provide useful guidance on errors for computationally expensive methods. We revisit earlier results \cite{Colgain:2023bge} based on graphical profile likelihoods to test how conclusions change with the full FC prescription. 

A secondary motivation is to employ the FC prescription in full generality in a controlled, computationally tractable setting. Following \cite{Gomez-Valent:2022hkb} (see also \cite{Trotta:2017wnx} where this possibility is raised), we bin the MCMC chain, thereby ensuring that one is working with the same underlying information in both Bayesian and frequentist settings. This sharpens any comparison. Beyond our simplified setting, the full-blown FC prescription has been employed in the literature \cite{SPIDER:2021ncy, LiteBIRD:2023zmo}. Nevertheless, one usually injects the same input cosmology and makes assumptions on the number of parameters where the confidence intervals depend on the number of parameters being fitted \cite{LiteBIRD:2023zmo} (see Fig. 6-8). 

\section{Methods}
The late Universe $\Lambda$CDM model with Hubble parameter, 
\begin{equation}
\label{eq:lcdm}
H(z) = H_0 \sqrt{1-\Omega_m + \Omega_m (1+z)^3}, 
\end{equation}
transitions from a 2D model with fitting parameters $(H_0, \Omega_m)$ into an effective 1D model with combined parameter $H_0 \sqrt{\Omega_m}$, since $H(z)$ scales as $H(z) \sim H_0 \sqrt{\Omega_m} (1+z)^{\frac{3}{2}}$ as one removes lower redshift observational Hubble data (OHD) in the late Universe \cite{Colgain:2022tql}. In MCMC analysis, one sees this as the gradual distortion of 2D confidence ellipses into banana-shaped contours - a hallmark of the  unbroken degeneracy between $(H_0, \Omega_m)$ - as low redshift data is removed. 

Given this obstacle to constraining $(H_0, \Omega_m)$ at exclusively high redshift, one may be puzzled for the motivation for removing low redshift data. However, if one wants to perform consistency checks \cite{Akarsu:2024qiq} of the flat $\Lambda$CDM model at different epochs (redshifts) in response to $\Lambda$CDM tensions \cite{DiValentino:2021izs, Perivolaropoulos:2021jda, Abdalla:2022yfr, CosmoVerse:2025txj}, in order to check that the fitting parameters are indeed constant and the $\Lambda$CDM model consistently fits the data at different epochs, one is forced to bin the data. What matters here is constancy of the parameters within the errors, thus motivating the focus on credible and confidence intervals that define the errors.    

\subsection{Data}
\begin{figure}[htb]
   \centering
\includegraphics[width=90mm]{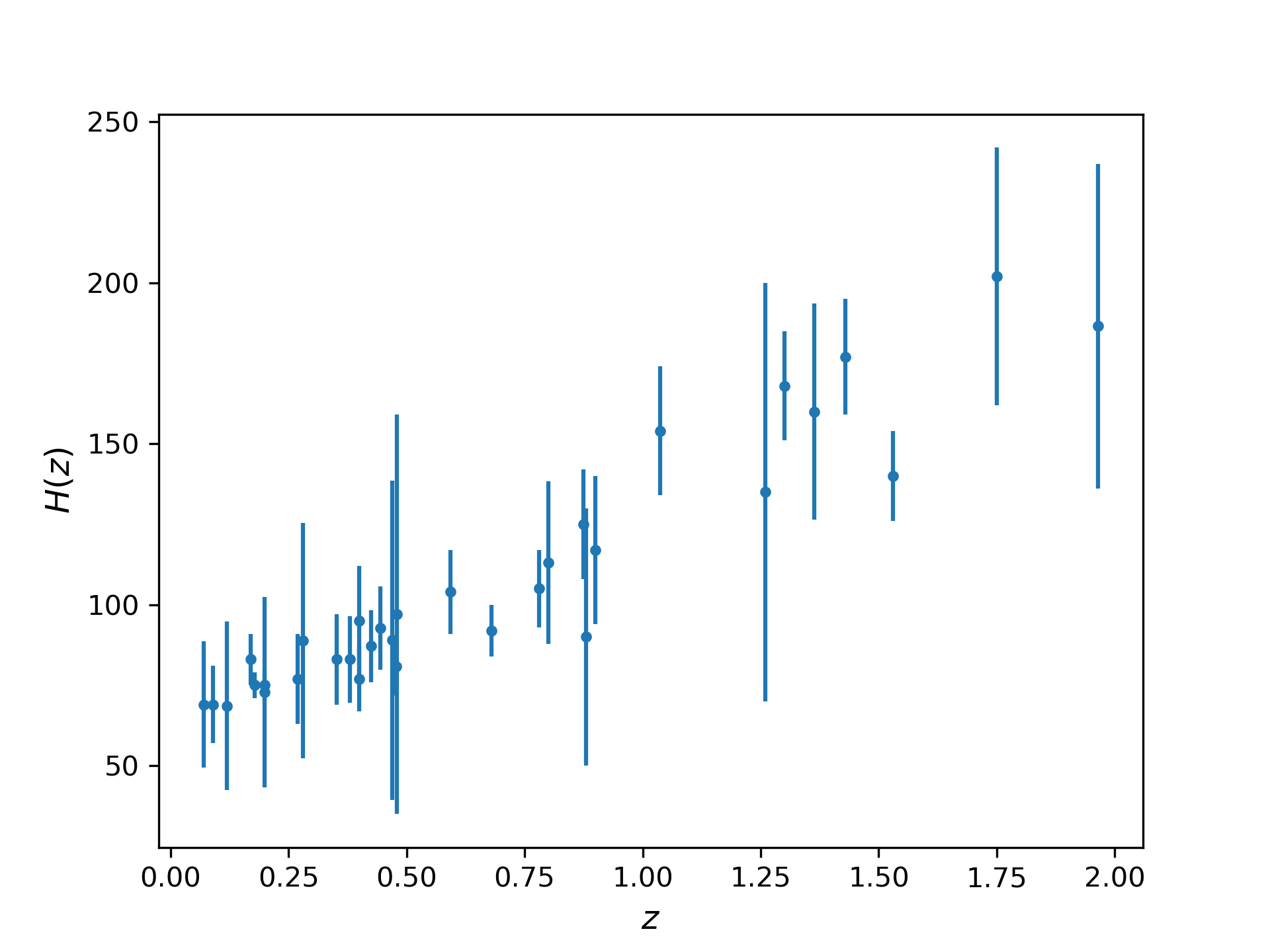} 
\caption{33 OHD constraints from the CC program.}
\label{fig:CCdata}
\end{figure}

In our analysis we make use of observed OHD from the cosmic chronometer research program \cite{Jimenez:2001gg, Moresco:2016mzx, Moresco:2023zys}. In particular we use the data from Table 1.1 of \cite{Moresco:2023zys} but remove two data points at $z=0.75$ that are not independent. This leaves 33 $H(z)$ constraints spanning the redshift range $ 0.07 \leq z \leq 1.965$. We illustrate the data in Fig. \ref{fig:CCdata}. As pointed out in \cite{Colgain:2023bge}, beyond $z \sim 1$, the higher redshift CC data prefers larger values of $H_0 \sim 150$ km/s/Mpc. It is clear from the raw data in Fig. \ref{fig:CCdata} that beyond $z \sim 1$, the errors are large and there is little variation in the data points, so one can almost fit a horizontal line through the error bars. This corresponds to an arguably unphysical situation where  physical matter density $\omega_m = \Omega_m h^2$ ($h := H_0/[100 \textrm{ km/s/Mpc}$]) is zero. Irrespective of the physical problem with $\omega_m \sim 0$, what interests us here is that there is evolution in the preferred $(H_0, \Omega_m)$ parameters as one bins the CC data, since the full CC sample prefers $H_0 \sim 70$ km/s/Mpc. This runs the risk that our findings may depend on the idiosyncrasies of the CC dataset. 

We do not incorporate the covariance matrix from \cite{Moresco:2020fbm} in our analysis. The reason being is that we are primarily only interested in the CC data as illustrative $H(z)$ data that allows us to access regimes in parameter space with non-Gaussian posteriors. This explains why we also analyse mock DESI $H(z)$ data in parallel. Nevertheless, we do have a second motivation. Ref. \cite{Colgain:2023bge} reported a $\sim 2 \sigma$ shift in $\Lambda$CDM parameters both using simulations and graphical profile likelihoods, where the analysis was performed without covariance matrix, and here will attempt to recover that $\sim 2 \sigma$ shift using the FC method. Restoring the covariance matrix in this work would prevent direct comparison with \cite{Colgain:2023bge}. We acknowledge that adding a covariance matrix simply inflates errors, thus making shifts in parameters less statistically significant, but the focus here is not the statistical significance of the shift but rather making contact with earlier results in the literature under similar assumptions.

\begin{figure}[htb]
   \centering
\includegraphics[width=90mm]{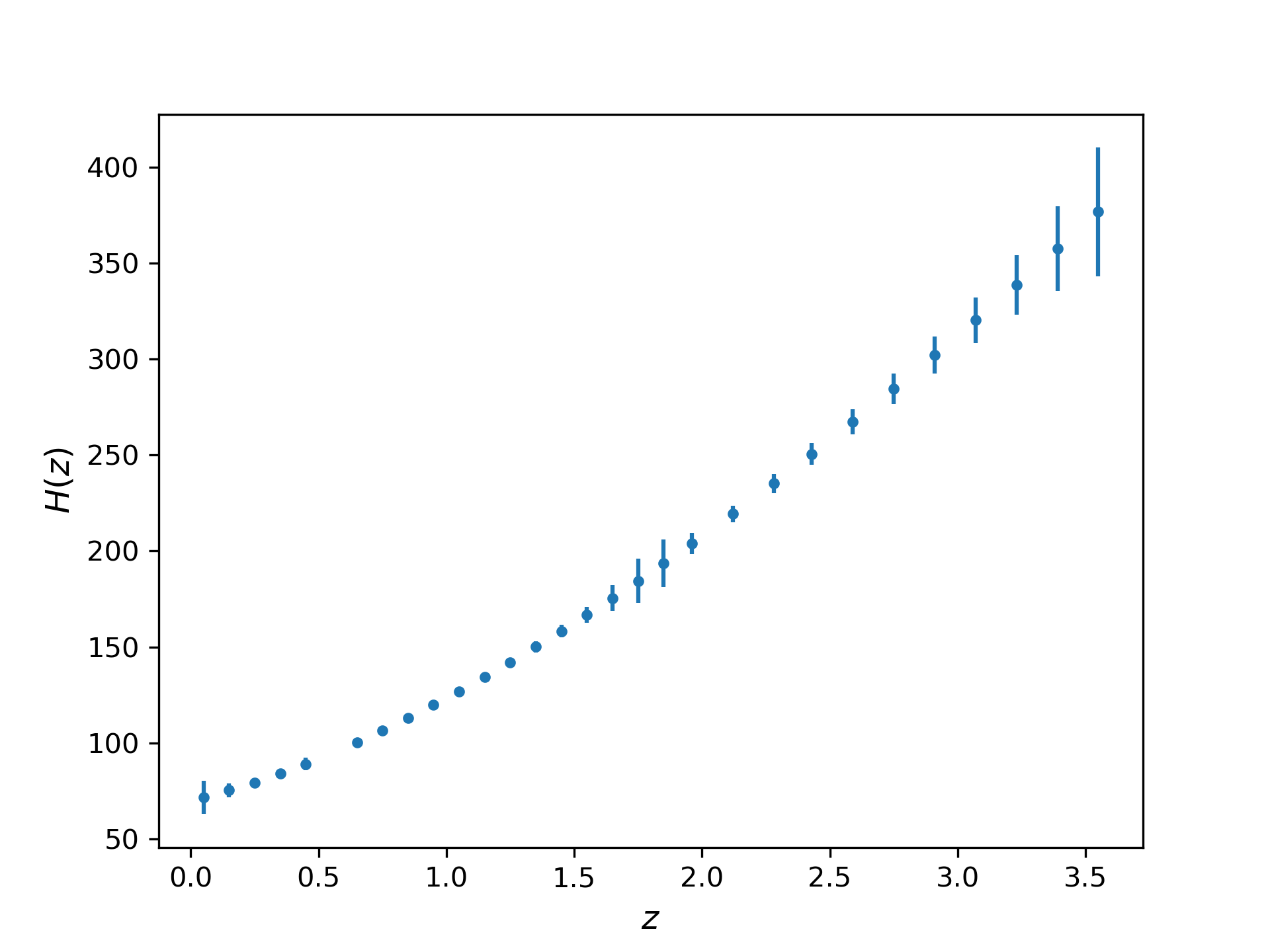} 
\caption{29 OHD constraints from DESI forecast with canonical $\Lambda$CDM cosmology.}
\label{fig:DESIdata}
\end{figure}

We can counter the concern of evolving $(H_0, \Omega_m)$ parameters in CC data by also studying mock data. For this reason, we make use of forecast DESI OHD constraints \cite{DESI:2016fyo}. The data comprises 29 data points with redshifts $0.05 \leq z \leq 3.55$ and percentage errors, but requires one to assume a cosmology. Concretely, we choose the canonical values $H_0 = 70$ km/s/Mpc and $\Omega_m = 0.3$ for the $\Lambda$CDM model. The resulting realisation of the data is presented in Fig. \ref{fig:DESIdata}. Given that we mock this dataset, it cannot be impacted by observational systematics, thereby making it more robust. Comparing Fig. \ref{fig:DESIdata} to Fig. \ref{fig:CCdata}, one notes that the fractional errors are considerably smaller in Fig. \ref{fig:DESIdata} and the data probes higher redshifts. To analyse the dataset, we follow \cite{Colgain:2022tql} and split it into four redshift bins. 

Note, we change our binning strategy between the two datasets. The motivation is straightforward. Fig. \ref{fig:CCdata} has large fractional errors, so the data poorly constrains the model when binned. For this reason, we work with a single bin, but we change the redshift range of the bin to remove lower redshift constraints. On the contrary, the data in Fig. \ref{fig:DESIdata} exhibit much smaller fractional errors, allowing us to consider more bins. The key point is that whatever conclusions we arrive at must be independent of the dataset and the binning strategy.  

\subsection{MCMC}
Throughout the analysis, we constantly compare back to Bayesian posteriors from  MCMC, where we make use of \textit{emcee} \cite{Foreman-Mackey:2012any}. To sharpen the comparison, we recycle the MCMC chain to construct frequentist confidence intervals, where we impose uniform priors $H_0 \in [0, 200]$ km/s/Mpc and $\Omega_m \in [0, 1]$. As explained by Trotta \cite{Trotta:2017wnx}, there is no obstacle to  constructing profile likelihoods directly from the MCMC chain; it is simply less efficient because MCMC is, unlike gradient descent, not designed for optimisation. However, given the simplicity of our setup, it is computationally inexpensive to run long MCMC chains well beyond the point of convergence to ensure good agreement with optimisation. We refer the reader to Fig. 4 of \cite{Colgain:2024clf} where it is confirmed that optimisation and binning the MCMC chain leads to good agreement even when the chain has converged in not necessarily a very long chain, and in a more complicated situation where additional nuisance parameters beyond $(H_0, \Omega_m)$ are present.  

\begin{figure}[htb]
   \centering
\includegraphics[width=80mm]{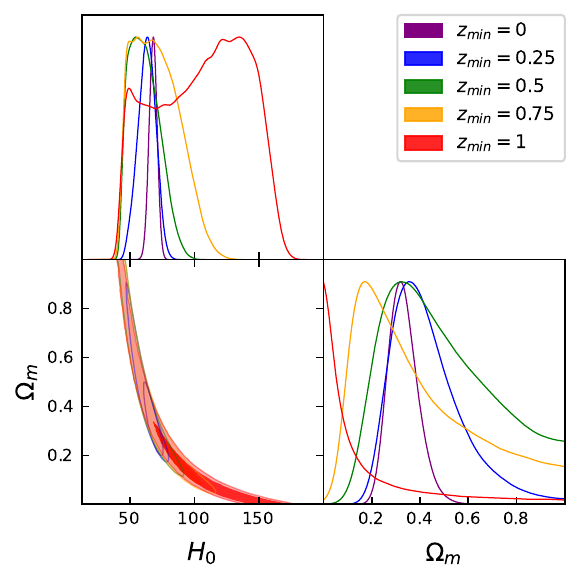} 
\caption{MCMC posteriors for CC OHD with 5 low redshift cuts $z > z_{\textrm{min}}$. 2D posteriors trace curves of constant $H_0 \sqrt{\Omega_m}$ at higher redshifts as this is the only parameter combination well constrained by the data. The marginalised 1D posteriors become steadily less Gaussian as effective redshift is increased. $H_0$ is in units of km/s/Mpc throughout this work.}
\label{fig:CC_posteriors}
\end{figure}

\begin{figure}[htb]
   \centering
\includegraphics[width=80mm]{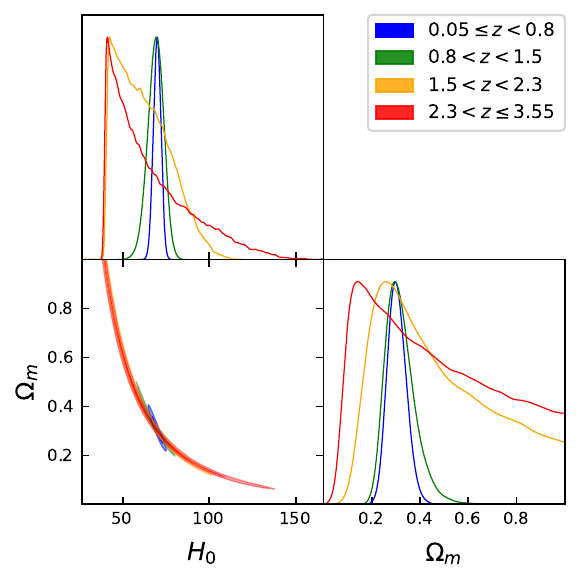} 
\caption{Same as Fig. \ref{fig:CC_posteriors} but for DESI forecast OHD in 4 redshift bins.}
\label{fig:DESI_posteriors}
\end{figure}

Given the MCMC chain, we process it in two different ways. The first is to extract a median, 16$^{\textrm{th}}$ and $84^{\textrm{th}}$ percentiles for each parameter $\theta \in \{H_0, \Omega_m \}$. This gives us Bayesian credible intervals.\footnote{Beyond Gaussian posteriors, care is required comparing to frequentist methods because i) the median need not track the mode, the equivalent of the frequentist MLE, and ii) ensuring equal probabilities in the tails through percentiles does not prioritise highest density regions. This  can be contrasted with the graphical profile likelihood methods we employ, so any comparison may not be like-for-like.} Before moving onto profile likelihoods, it is instructive to confirm that restricting OHD to higher redshifts leads to non-Gaussian posteriors. In Fig. \ref{fig:CC_posteriors} and Fig.  \ref{fig:DESI_posteriors} we show $H_0$ and $\Omega_m$ posteriors from representative MCMC chains for observed CC and mock DESI data, respectively. We remind the reader that we have injected $H_0 = 70$ km/s/Mpc and $\Omega_m = 0.3$ for the mock data. Irrespective of the dataset, what happens is that as one increases the effective redshift, Gaussian confidence ellipses in the $(H_0, \Omega_m)$-plane in lower redshift bins get stretched into contours of constant $H_0 \sqrt{\Omega_m}$ in higher redshift bins. These contours can stretch as far as the $\Omega_m$ boundaries, so all inferences become prior dependent. 

In addition, projection effects are evident in the posteriors. The secondary peak in the $z_{\textrm{min}}=1$ $H_0$ posterior in Fig. \ref{fig:CC_posteriors} comes from projecting MCMC configurations in the top left corner of the 2D posterior that differ appreciably in $\Omega_m$ values, but differ only marginally in $H_0$ values. Only the $z_{\textrm{min}} = 0$ and $z_{\textrm{min}} = 0.25$ posteriors are anyway close to the Gaussian regime. In Fig. \ref{fig:DESI_posteriors} projection effects are evident in the fact that the peaks of the 1D posteriors differ from the injected parameters. Once again, only for the 2 lowest redshift bins are we close to the Gaussian regime. A comparison of Fig. \ref{fig:CC_posteriors} and Fig. \ref{fig:DESI_posteriors} reveals the red posterior in Fig. \ref{fig:CC_posteriors} as the only one where $\Omega_m = 0$ is probable. As explained earlier, this can be attributed to a idiosyncrasy or quirk in the observed CC data. However, in general, high redshift OHD constrains the combination $H_0 \sqrt{\Omega_m}$ to a non-zero value, and provided $H_0$ does not run away to infinity, $\Omega_m$ cannot be zero. The main takeaway is that while the $\Omega_m \leq 1$ bound impacts the posteriors, the $\Omega_m \geq 0$ bound generically does not impact our posteriors unless physical matter density is zero.

As explained above, we will recycle the MCMC chain to reconstruct our profile likelihood confidence intervals. What this means is that when we divide the chain in narrow $H_0$ or $\Omega_m$ bins, the number of MCMC configurations and corresponding mock simulations in each bin varies according to the projected 1D posterior. This will lead to some noise in our profile likelihoods for points in parameter space that the MCMC algorithm has poorly explored. For this reason, all our profile likelihoods are viewed through the prism of the MCMC chain.

\subsection{Profile Likelihoods}
To construct our graphical profile likelihoods we reycle the MCMC chain \cite{Trotta:2017wnx, Gomez-Valent:2022hkb}, thereby providing the sharpest comparison to Bayesian methods. Concretely, we identify the range of the parameter of interest $\theta \in \{H_0, \Omega_m \}$ from the MCMC chain, i. e. the maximum and minimum value, and divide the range up into approximately 200 evenly spaced bins. We label each bin by the value of the parameter at the centre. What this means in practice is that if $\theta = H_0$, then the $H_0$ values in each bin span a narrow range, but the $\Omega_m$ values may span a much greater range. 

Once one bins the MCMC chain as outlined in the previous section, graphical profile likelihoods \cite{Herold:2024enb} are constructed by identifying the minimum value of the $\chi_{\textrm{min}}^2 (\theta)$ in each $\theta$-bin. Note when $\theta = H_0$, this corresponds to finding the minimum of the $\chi^2$ for approximately fixed $H_0$ values, more accurately $H_0$ values in a narrow bin, for any value of $\Omega_m$ subject to our bounds, and \textit{vice versa}. The profile likelihood ratio is then \cite{Trotta:2017wnx, Gomez-Valent:2022hkb} 
\begin{equation}
\label{eq:R}
    R(\theta) = \exp \left(- \frac{1}{2} \Delta \chi^2 \right) =  \exp \left(- \frac{1}{2} (\chi^2_{\textrm{min}}(\theta) - \chi_{\textrm{min}}^2 ) \right), 
\end{equation}
where $\chi^2_{\textrm{min}}$ without an argument denotes the global minimum of the $\chi^2$ from the MCMC chain. This global minimum occurs in one of the $\theta$-bins, so the profile likelihood ratio peaks at $R(\theta) = 1$ by construction. Since MCMC is a stochastic process, the binned MCMC chain leads to a different number of MCMC configurations in each bin. In particular, far from the median values for the chain, the bins will be sparsely populated. Nevertheless, provided there is more than one configuration in each bin, we include the bin centre value in our profile likelihood ratio (\ref{eq:R}). The consequence of having so few configurations in a bin is that $\chi^2_{\textrm{min}}(\theta)$ may be larger than the minimum $\chi^2$ one would get from gradient descent (optimisation). This leads to relatively lower values of $R(\theta)$. One can see from Fig. 4 of \cite{Colgain:2024clf} that the $R(\theta)$ values one gets from MCMC converge to those one gets from optimisation from below, which is the expected outcome. Obviously, in the tails of the profile likelihoods where $R(\theta)$ is small anyway, such differences are negligible.  

\begin{figure}[htb]
   \centering
\includegraphics[width=90mm]{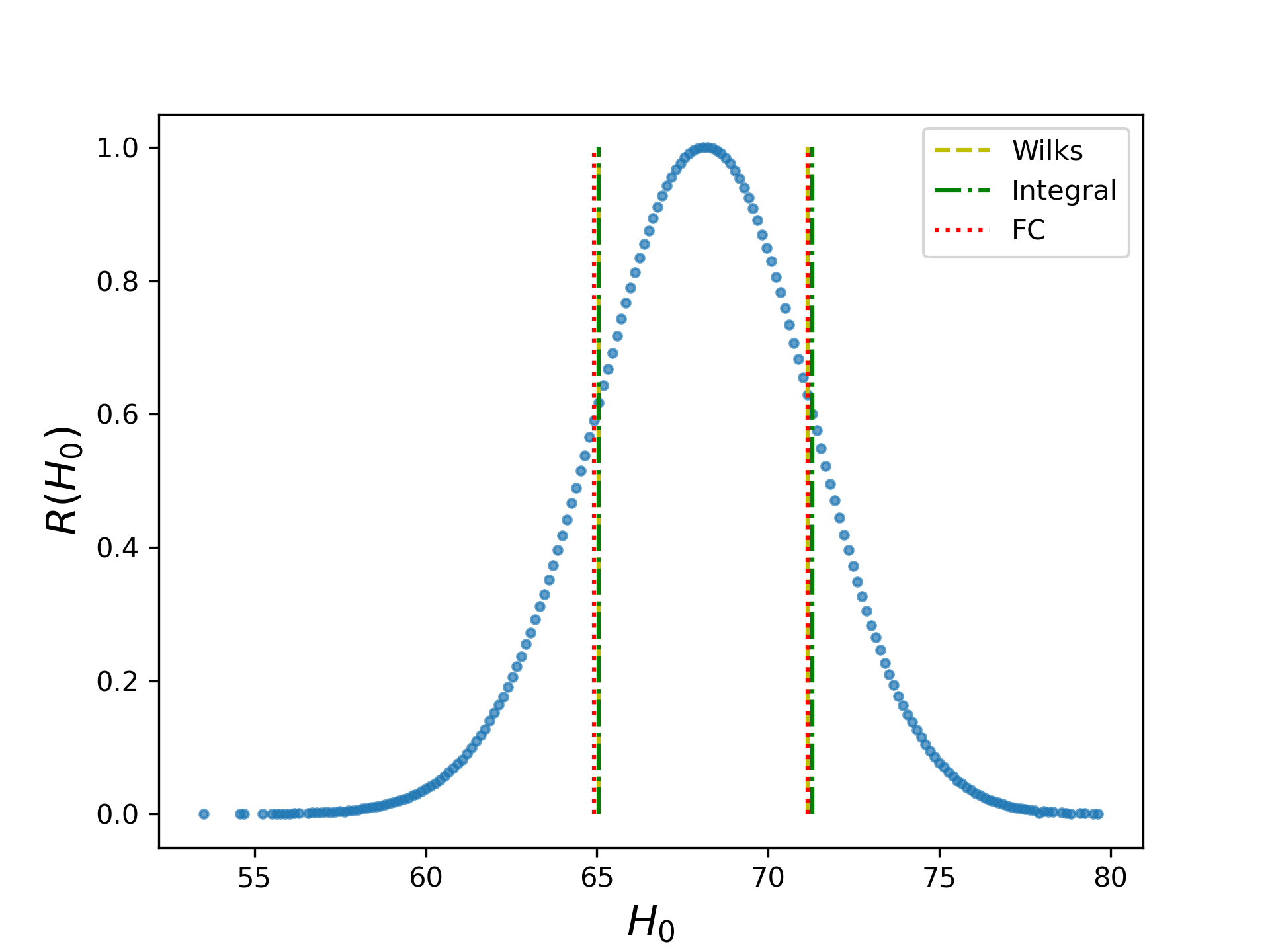} 
\caption{Graphical profile likelihood for the $H_0$ parameter and CC data with $z_{\textrm{min}} = 0$ (full sample). The $68\%$ confidence intervals for three methods are denoted by vertical lines.}
\label{fig:h0_zmin00_profile}
\end{figure}

In Fig. \ref{fig:h0_zmin00_profile} we present the $H_0$ graphical profile likelihood for CC data with $z > z_{\textrm{min}} = 0$ corresponding to the full sample. The profile likelihood is visibly close to Gaussian, but the tails towards lower $H_0$ values are noticeably longer, so $R(H_0)$ cannot be exactly Gaussian. Moreover, in the tails, it is evident that there are bins without a configuration that have been omitted. It is also worth noting that different methods agree on the $68 \%$ confidence interval, providing an important consistency check of our methods in a regime that is as close to Gaussian as the CC data allows. We now turn our attention to explaining these confidence interval methods.  

\subsection{Wilks' theorem}
\label{sec:wilks}
Once we have constructed the graphical profile likelihood in equation (\ref{eq:R}), it is easy to identify regions corresponding to $\Delta \chi^2 \leq 1$ through $R(\theta) \gtrsim 0.607 $. From Fig. \ref{fig:h0_zmin00_profile} and later plots it should also be easy to draw a horizontal line by eye across from $R(\theta) \approx 0.6$ to check that Wilks' $68 \%$ confidence interval intersect the profile likelihood curve in the expected range. The reader should note that Wilks' theorem only strictly holds in the large sample limit \cite{Wilks}, where Bayesian posteriors and frequentist profile likelihoods are close to Gaussian \footnote{In cosmology it is easy to find small samples that give Gaussian constraints. A single anisotropic BAO constraint on $H(z)$ and $D_{M}(z) \propto \int 1/H(z) \, \dd z$ confronted to the flat $\Lambda$CDM model cleanly breaks the degeneracy between $(H_0, \Omega_m)$ and places one in the close to Gaussian regime \cite{Colgain:2022tql}.}. For this reason, we expect that the $\Delta \chi^2 \leq 1$ prescription breaks down as the posterior/profile likelihood becomes steadily more non-Gaussian (see Fig. \ref{fig:CC_posteriors} and Fig. \ref{fig:DESI_posteriors}). We remark that $\Delta \chi^2 \leq 1$ only requires information of the profile likelihood in the local vicinity of the MLE and does not make use of the global profile likelihood. As one proceeds to larger confidence intervals, e. g. $95 \%$ confidence level, corresponding to $\Delta \chi^2 \leq 4$, one clearly explores more of the profile likelihood. 

\subsection{Integration}
\label{sec:integration}
Given that Wilks' theorem builds a picture of the graphical profile likelihood \textit{locally} from the MLE outwards, one can envision using all the information in the profile likelihood. In \cite{Gomez-Valent:2022hkb} it was suggested that one could construct confidence intervals by integrating the \textit{normalised} profile likelihood, whereby the motivation was to identify projection effects in MCMC. Here, we take this proposal at face value and add it to our comparisons, since for close to Gaussian profile likelihoods, this is guaranteed to agree with Wilks' theorem. See section III of \cite{Colgain:2024clf} for further explanations of this point. That being said, the methods need not agree in the non-Gaussian regime. Moreover, this method becomes prior dependent whenever the profile likelihood is cut off by our MCMC bounds. 

The basic idea is to normalise $R(\theta)$ in equation (\ref{eq:R}) by the area under the profile likelihood curve \cite{Gomez-Valent:2022hkb}: 
\begin{equation}
    \label{eq:w}
    w(\theta) = \frac{R(\theta)}{\int R(\theta) \, \dd \theta}, 
\end{equation}
and identify the widest interval $[\theta^{(1)}, \theta^{(2)}]$ so that the following inequality holds: 
\begin{equation}
    \label{eq:w_solve}
    \int_{\theta^{(1)}}^{\theta^{(2)}} w(\theta) \, \dd \theta > 0.68, \quad w(\theta^{(1)}) = w (\theta^{(2)}).  
\end{equation}
This gives one a $68\%$ confidence interval. Since we bin the MCMC chain, which in turn gives rise to discretised profile likelihood ratios $R(\theta)$ (see Fig. \ref{fig:h0_zmin00_profile}), the last step is most easily executed through Simpson's rule for integration and a threshold $\kappa$ \cite{Colgain:2023bge}. More precisely, one defines $F_{\kappa} := \int_{R \geq \kappa} R(\theta) \, \dd \theta$. Since $R(\theta)$ is peaked at $R(\theta) = 1$ by construction in equation (\ref{eq:R}), one has $F_{\kappa=1} = 0$ and $F_{\kappa = 0} := F_0 = \int R(\theta) \, \dd \theta$. To satisfy (\ref{eq:w_solve}) one simply identifies the lowest value of $\kappa$ so that $F_{\kappa}/F_0 > 0.68$. It is worth noting that in both these methods the values of $R(\theta)$ at the ends of the confidence interval agree.  

\subsection{Feldman-Cousins}
Both Wilks' theorem and profile likelihood integration only make use of a single realisation of the observed data. Ref. \cite{Feldman:1997qc} extends the earlier Neyman confidence interval construction \cite{Neyman:1937uhy} to introduce an ordering that addresses parameters with boundaries protecting one from unphysical regimes. In particular, Feldman \& Cousins study one measured quantity with one unknown parameter, including a Gaussian with a boundary \cite{Feldman:1997qc}. In the Gaussian setting, it can be shown that a boundary always contracts Wilks' confidence intervals \cite{Colgain:2024clf}, so FC confidence intervals are bounded above by Wilks' confidence intervals. We will be interested in seeing if there are generic hierarchies between the different frequentist methods.  

We begin by presenting an overview of the FC method. To construct FC confidence intervals, one defines a likelihood-ratio ordering principle for each possible value of the parameter of interest $\theta$:
\begin{equation}
\label{eq:FC_ratio}
R(x;\theta) = \frac{\mathcal{L}(x|\theta)}{\mathcal{L}(x|\hat{\theta})},
\end{equation}
where $\mathcal{L}(x|\theta)$ is the likelihood of observing data $x$ for parameter value $\theta$, and $\hat{\theta}$ is the value that maximises the likelihood for that same data $x$. For each $\theta$, the probability distribution function of the data $p(x|\theta)$ is used to define an acceptance region $\mathcal{A}(\theta)$ by including outcomes with the largest $R(x;\theta)$ until
\begin{equation}
\int_{\mathcal{A}(\theta)} p(x|\theta)\, \dd x = 1-\alpha.
\end{equation}
The FC confidence interval is then given by the set of all $\theta$ values for which the observed data $x$ falls inside $\mathcal{A}(\theta)$. This procedure ensures proper frequentist coverage while avoiding the need to predetermine an ordering rule.
An important point to note is that for continuous observables, the FC ordering principle implies $R(x_1|\theta) = R(x_2|\theta)$ at the extremities of the acceptance intervals. However, when we consider discrete observables (such as a Poisson distributed observable), this is not necessarily the case.

In our construction of FC confidence intervals we make use of the binned MCMC chain, so the starting point of the analysis is the same as the graphical profile likelihood, namely 200-odd evenly spaced bins in a parameter $\theta \in \{H_0, \Omega_m \}$ labelled by the centre of the bin. However, in contrast to the profile likelihood construction, we remove all bins with less than 100 $(H_0, \Omega_m)$ values; some of the bins have many multiples of 100 values, but at the extremities of the $\theta$ range, the numbers decrease. The reason we keep at least 100 values is that we have to identify percentiles later. From this point, our analysis largely follows the methodology of ref. \cite{LiteBIRD:2023zmo}, but there are key differences that we comment on below. Each MCMC configuration in a bin serves as the input parameters for a single mock, whereby we move all the $H(z)$ constraints in Fig. \ref{fig:CCdata} and Fig. \ref{fig:DESIdata} in the relevant redshift range to the $H(z)$ of the input $\Lambda$CDM cosmology, before selecting new data points randomly using standard deviations that coincide with the error bars of the $H(z)$ constraint. 

For each mock realisation of the data constructed in this manner, we fit back the $\Lambda$CDM parameters $(H_0, \Omega_m)$ subject to the constraints $H_0 \in [0, 200]$ and $\Omega_m \in [0, 1]$. These bounds are the same as the uniform priors in the MCMC analysis and we comment on relaxing them later. Focusing on one of the parameters $\theta \in \{H_0, \Omega_m \}$ and its best fit $\hat{\theta}$, while the other becomes an auxiliary parameter, we then evaluate the likelihood ratio (\ref{eq:FC_ratio}).  
Tailoring to our MCMC binned analysis, $\theta$ denotes the value of the parameter at the centre of the bin of the injected cosmology, and $\hat{\theta}$ is the bin centre value that maximises the likelihood. Throughout, $R(x; \theta)$ is evaluated at the best fit auxiliary parameter. We comment on this further below. 

\begin{figure}[htb]
   \centering
\includegraphics[width=90mm]{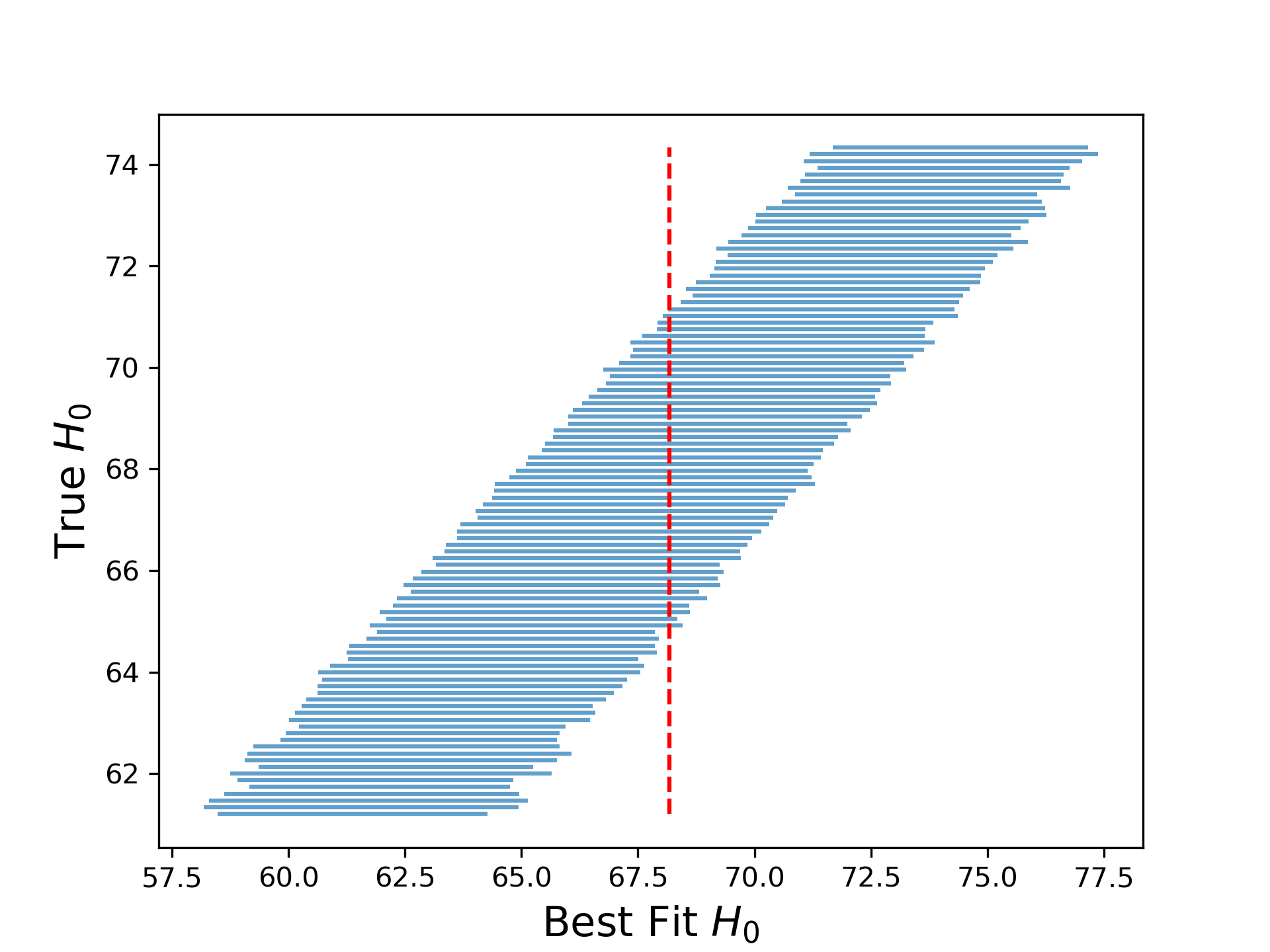} 
\caption{The $68 \%$ confidence belt for the $H_0$ parameter with $z_{\textrm{min}} = 0$ (full sample). Horizontal lines denote acceptance intervals and the red dashed line is the observed value. The ends of the belt draw visibly parallel lines characteristic of a Gaussian distribution.}
\label{fig:h0_zmin00_belt}
\end{figure}

For each simulation $x$, one has the output of the best fit or true value of the parameter of interest $\hat{\theta}$ and a likelihood ratio $R(x; \theta)$, $(\hat{\theta}, R(x; \theta))$. Collecting all of these pairs one isolates all the $\hat{\theta}$ in each bin corresponding to the $68 \%$ of mock simulations with the largest $R$. This leads to the horizontal lines corresponding to acceptance intervals in Fig. \ref{fig:h0_zmin00_belt}, where we have chosen $\theta= H_0$. True $H_0$ denotes the bin centre $H_0$ and Best Fit $H_0$ allows us both to track the range of $H_0$ MLEs in the $68 \%$ acceptance intervals, but also allows us to pinpoint the location of the observed value, namely the best fit $H_0$ from the observed data, which we denote with a vertical dashed red line. Finally, the $68 \%$ confidence interval is the union of all True $H_0$ values corresponding to an acceptance interval that intersects the vertical line. 
As pointed out in Fig. 3 of the FC paper \cite{Feldman:1997qc}, in the absence of boundaries, Gaussian distributions give rise to a confidence belt where the extremities of the acceptance intervals trace parallel lines. Here, Fig. \ref{fig:h0_zmin00_belt} is constructed from the same MCMC chain as Fig. \ref{fig:h0_zmin00_profile}, where the latter is visibly close to Gaussian, so one expects parallel lines, and this is what we see. 

In contrast to particle physics \cite{Cousins:1991qz, Feldman:1997qc, NOvA:2022wnj}, in cosmology the parameter and measured quantity are usually the same parameter, e. g. the tensor-to-scalar ratio $r_*$ from \cite{SPIDER:2021ncy, LiteBIRD:2023zmo}. Thus, here we have two parameters $(H_0, \Omega_m)$, but also two measured quantities $(H_0, \Omega_m)$. As is common in the cosmology literature \cite{SPIDER:2021ncy, LiteBIRD:2023zmo}, we isolate a parameter of interest $\theta \in \{ H_0, \Omega_m \}$ and treat the auxiliary parameter as a nuisance parameter. As explained in \cite{NOvA:2022wnj}, one can hold nuisance parameters fixed to their \textit{a priori} assumed values when one generates mock simulations. This strategy is employed in \cite{SPIDER:2021ncy, LiteBIRD:2023zmo}. The problem with this approach is that the best fit value of $H_0$ to a mock realisation changes depending on whether one fixes or fits $\Omega_m$. Moreover, since the auxiliary parameter $\Omega_m$ appears in the likelihood $\mathcal{L} (x |\theta)$ in equation (\ref{eq:FC_ratio}), fixing $\Omega_m$ impacts $\mathcal{L} (x | \hat{\theta})$, running the risk that the ratio $R(x; \theta)$ changes. Evidently, both the best fits in the acceptance intervals and likelihood ratios can be impacted by fixing parameters. Ultimately, it is better to fit nuisance parameters in order to ensure that they extract as much information as possible from the mocks.

\subsection{Summary}
Having introduced various methods to construct frequentist confidence intervals starting from binned MCMC chains, we comment on their feasibility in cosmological settings. To begin, Bayesian MCMC analysis is par for the course in cosmology. From there, it is relatively straightforward to bin the MCMC chain and employ Wilks' theorem or integrate the normalised graphical profile likelihood ratio (Fig. \ref{fig:h0_zmin00_profile}). The advantage of the former is that one only needs information of $\Delta \chi^2$ in the vicinity of the MLE, but the downside is that the theorem only holds in the large sample limit where posteriors and profile likelihoods are Gaussian. While it is routinely assumed in the literature that one is in the Gaussian regime, it is prudent to check \cite{Herold:2024enb}. Integrating the profile likelihood is computationally more expensive requiring a global reconstruction of $R(\theta)$ through equation (\ref{eq:R}) before extracting an interval from equation (\ref{eq:w_solve}). Nevertheless, nothing in the method is specific to Gaussian distributions, so potentially it may be more general, while guaranteeing the same results in the Gaussian regime. That being said, the method is not based on a theorem. Finally, one can embrace the FC prescription, where simulations may be computationally expensive, requiring further simplifying assumptions. 

Here we point the reader to the analysis in \cite{LiteBIRD:2023zmo}, where the authors inject the same cosmology throughout with no variation in the mock input parameters and make choices about how many parameters to fit back to the mocks. As is clear from the analysis, fitting back a single parameter and fitting back three parameters can greatly inflate the confidence interval from the belt construction \cite{LiteBIRD:2023zmo}. These choices ultimately lead to a degree of arbitrariness in the process as the final confidence intervals depend on one's assumptions. It is worth stressing that our simplified setting here, OHD with two parameters and an MCMC chain, allows us to be completely general, allowing for randomness both in the mocking and fitting process. In the appendix we check that fixing one of the parameters contracts confidence intervals and restores Gaussian distributions despite one being generically in a non-Gaussian regime. Moreover, even in the Gaussian regime, we find that fixing a parameter contracts the confidence intervals. Taken together with ref. \cite{LiteBIRD:2023zmo}, this cautions that one should not fix parameters when fitting the mock simulations integral to the Neyman and FC methods. 

We end this section with a necessary comment on boundaries. The relevant boundaries come from the traditional $\Lambda$CDM $\Omega_m \in [0, 1]$ prior, but all posteriors are unrestricted by a $H_0 \in [0, 200]$ km/s/Mpc prior. One can in principle try to relax the $\Omega_m \leq 1$ bound by allowing the cosmological constant $\Lambda \propto (1-\Omega_m)$ to change sign, but allowing $\Omega_m < 0$ is mathematically problematic as one encounters a minus sign in the square root. To see this, recall $H(z)$ from equation (\ref{eq:lcdm}). At lower redshifts $z \ll 1$, the dark energy term $1-\Omega_m$ may prevent a negative sign in the square root when $\Omega_m < 0$, however since only $\Omega_m (1+z)^3$ grows with redshift, negative signs in the square root are difficult to avoid at higher redshifts when $\Omega_m < 0$. We have checked that allowing mock simulation best fit parameters into the $\Omega_m > 1$ regime does not change our results. The reason this is the case is that all of our $\Omega_m$ MLEs in observed/mock data occur in the range $\Omega_m \lesssim 0.45$ (see later Table \ref{tab:CC} and Table \ref{tab:DESI}), thus ensuring that the $\Omega_m$ boundary is approximately $2 \sigma$ removed. In the Gaussian setting, this is far enough removed that the boundary has little impact \cite{Feldman:1997qc, Colgain:2024clf}. In contrast, we cannot relax $\Omega_m \geq 0$ for the best fit parameters without encountering problems with the square root.   

\section{Results}
\begin{figure}[htb]
   \centering
\includegraphics[width=85mm]{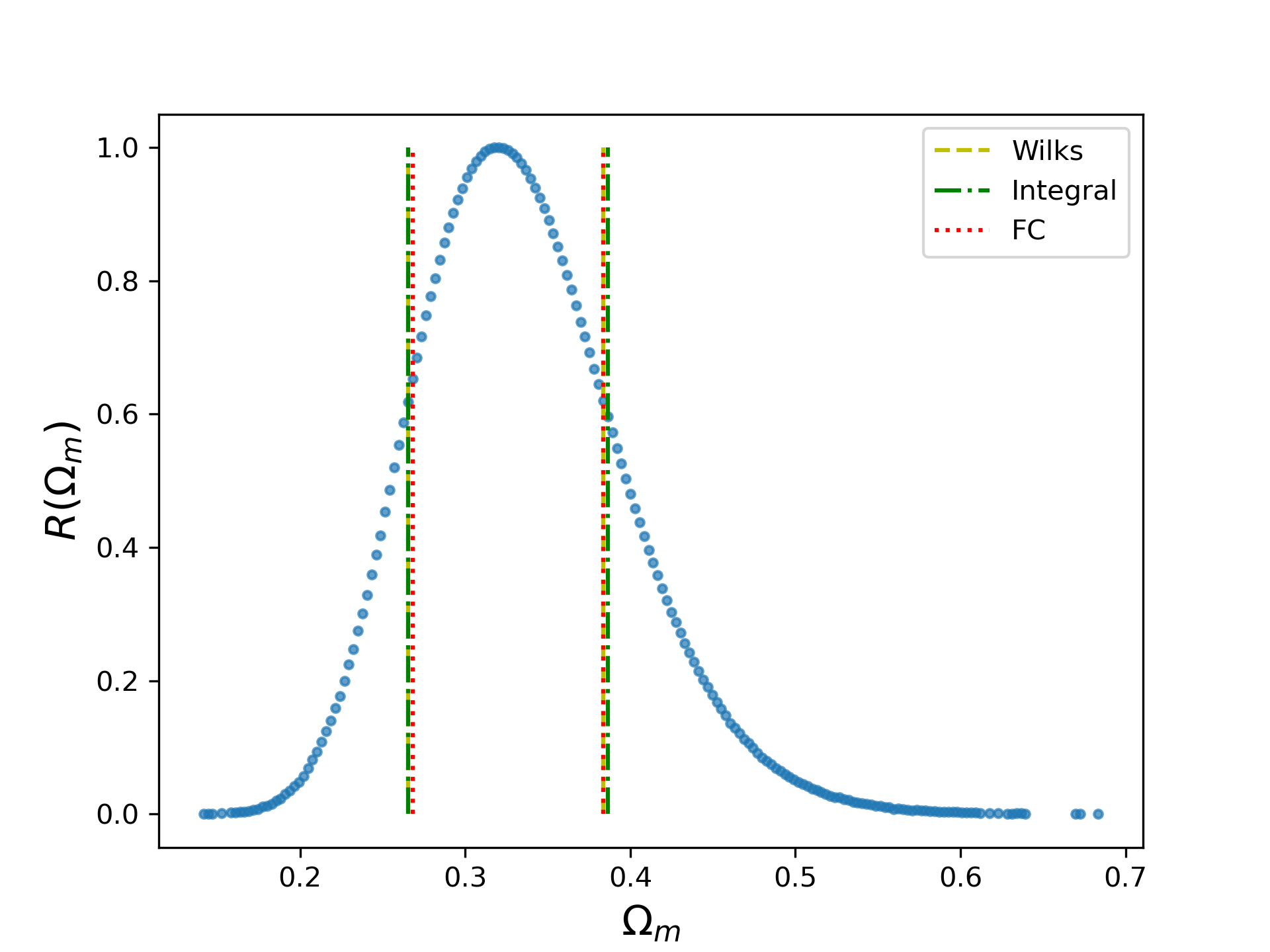} 
\caption{Graphical profile likelihood for the $\Omega_m$ parameter for the CC data with $z_{\textrm{min}} = 0$ (full sample). The $68\%$ confidence intervals for thr three methods are denoted by vertical lines and show excellent agreement.}
\label{fig:om_zmin00_profile}
\end{figure}

\begin{figure}[htb]
   \centering
\includegraphics[width=85mm]{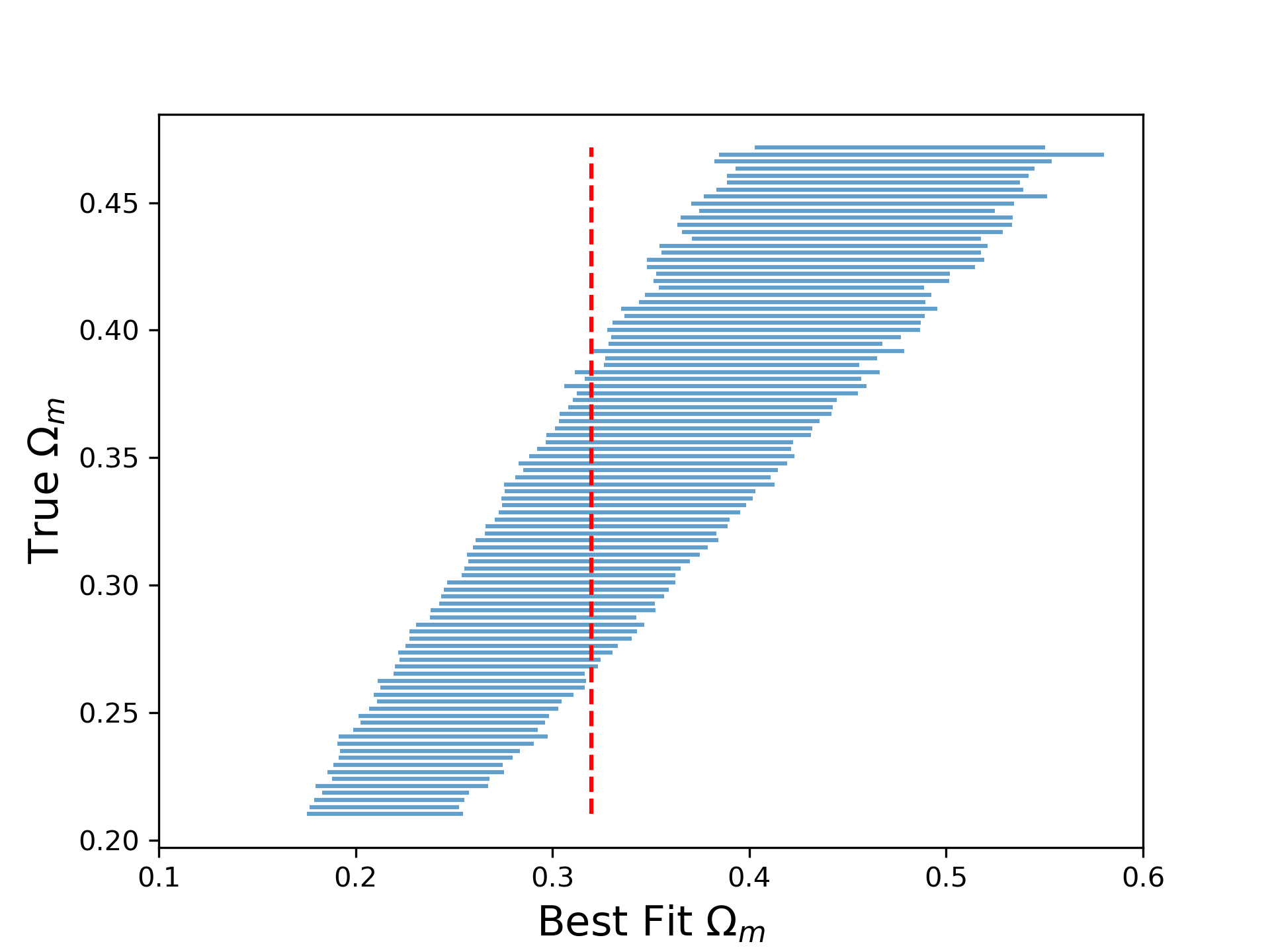} 
\caption{The $68 \%$ confidence belt for the $\Omega_m$ parameter and CC data with $z_{\textrm{min}} = 0$ (full sample).}
\label{fig:om_zmin00_belt}
\end{figure}

While the transition from Gaussian to non-Gaussian posteriors with increasing effective redshift is evident in all posteriors in Fig. \ref{fig:CC_posteriors} and Fig. \ref{fig:DESI_posteriors}, it is most pronounced in the projected 1D $\Omega_m$ posteriors. At a basic level, we can expect $\Omega_m$ graphical profile likelihoods $R(\Omega_m)$ to exhibit similar curves. See \cite{Colgain:2024ksa} for a realisation of this feature with DES SNe \cite{DES:2024jxu}. In Fig. \ref{fig:om_zmin00_profile} we present $R(\Omega_m)$ for the full CC dataset with $z_{\textrm{min}} = 0$. This profile is the $\Omega_m$ counterpart to Fig. \ref{fig:h0_zmin00_profile}. While the reader can confirm that all methods agree on $68 \%$ confidence intervals, a key point we want to get across is whatever mild non-Gaussian behavior is evident in the lopsided tails in Fig. \ref{fig:h0_zmin00_profile} is a little more pronounced in Fig. \ref{fig:om_zmin00_profile}. This more pronounced departure from exact Gaussianity is also evident in the confidence belt in the FC method, where we see from Fig. \ref{fig:om_zmin00_belt} that it no longer draws parallel lines, but the lines diverge. It is also worth noting that the acceptance intervals (horizontal lines) in the confidence belt do not probe as wide a range of True $\Omega_m$ values as the $R(\Omega_m)$ graphical profile likelihood as they are subject to different cutoffs on the number of MCMC configurations in the bin, i. e. 100 versus 1. That being said, what matters here is the range of True $\Omega_m$ values for which the observed MLE value for $\Omega_m$ intersects the acceptance intervals. 

\begin{figure}[htb]
   \centering
\includegraphics[width=85mm]{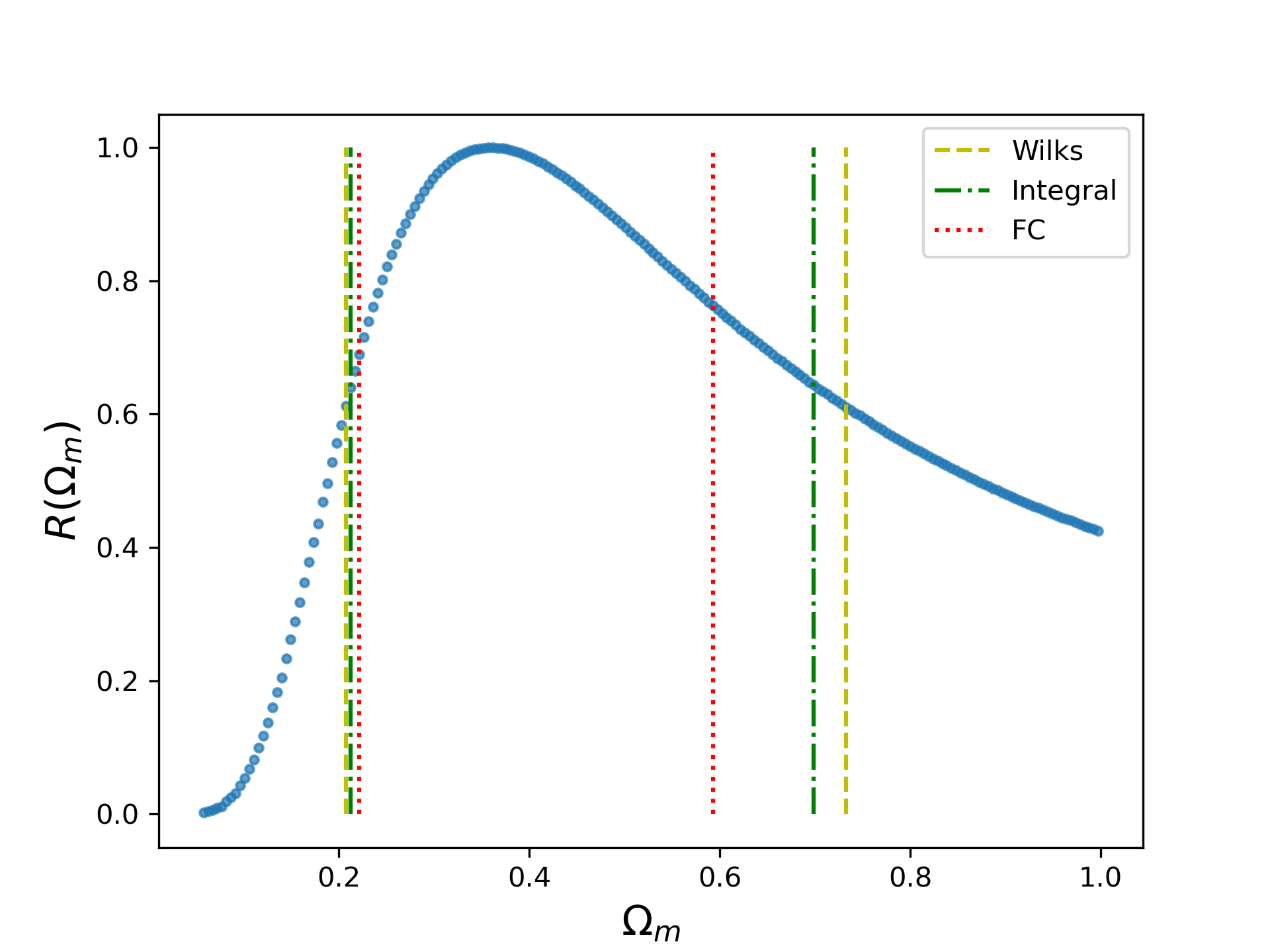} 
\caption{Same as Fig. \ref{fig:om_zmin00_profile} but for the CC data with $z_{\textrm{min}} = 0.5$.}
\label{fig:om_zmin05_profile}
\end{figure}

Now that we have recovered reasonable results with mildly non-Gaussian graphical profile likelihoods and FC confidence belts, we steadily increase the low redshift cutoff $z_{\textrm{min}}$ on the CC data. What we will be interested in seeing is that the $\Omega_m$ profile likelihoods trace qualitatively similar curves to the $\Omega_m$ posteriors in Fig. \ref{fig:CC_posteriors}. Note, there is no reason for these to be the same curve as the mode of the $\Omega_m$ posterior and the $\Omega_m$ MLE corresponding to the peak of the graphical profile likelihood can differ. Beginning with $z_{\textrm{min}} = 0.5$, we see that the $\Omega_m$ posterior is impacted by the $\Omega_m \leq 1$ bound. The corresponding graphical profile likelihood can be found in Fig. \ref{fig:om_zmin05_profile}. As the reader can confirm, the profile likelihood traces a qualitatively similar curve to the green $z_{\textrm{min}} = 0.5$ $\Omega_m$ posterior in Fig. \ref{fig:CC_posteriors}. 

\begin{figure}[htb]
   \centering
\includegraphics[width=85mm]{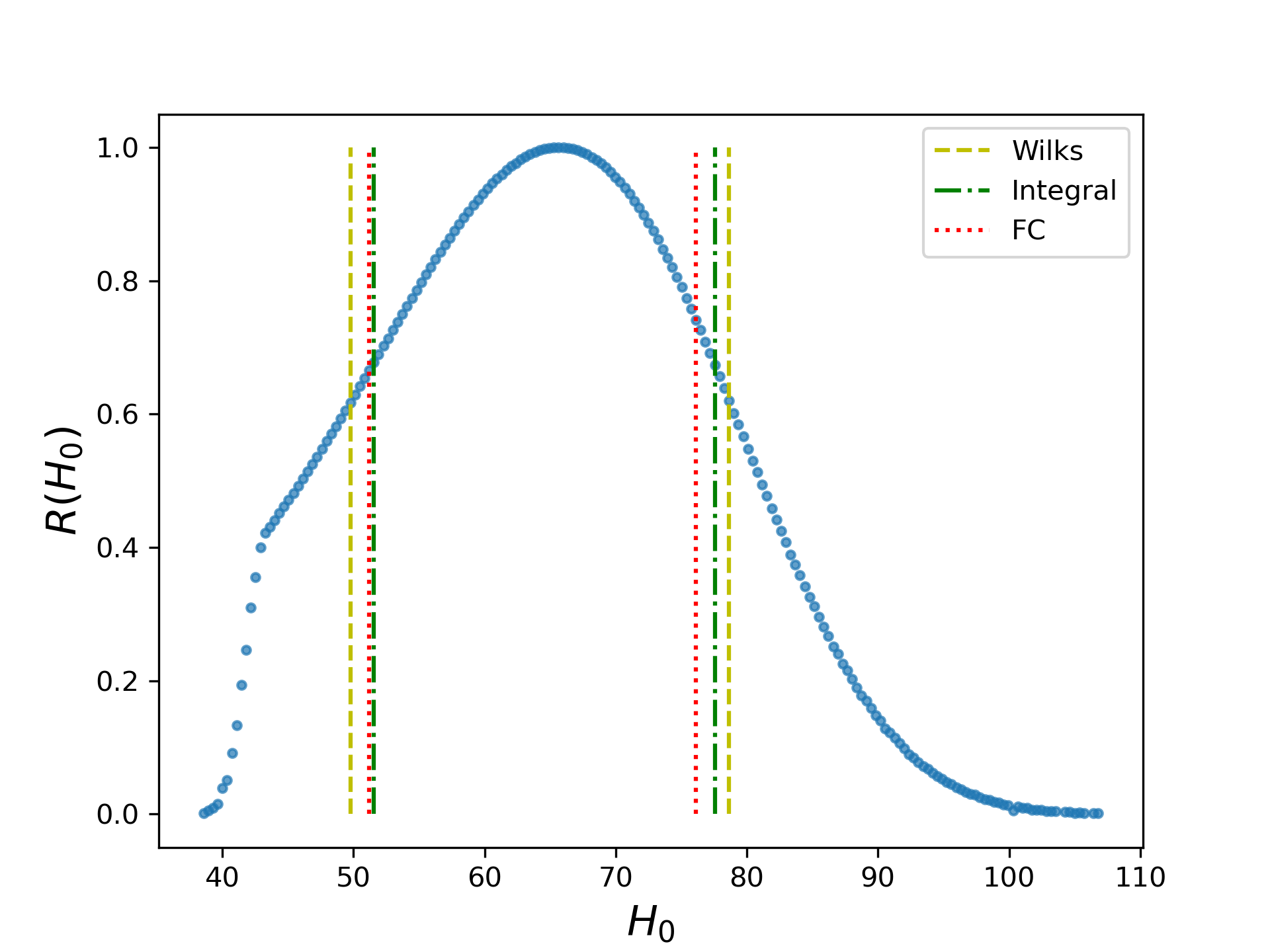} 
\caption{Same as Fig. \ref{fig:h0_zmin00_profile} but for the CC data with $z_{\textrm{min}} = 0.5$.}
\label{fig:oh0_zmin05_profile}
\end{figure}

We also note that while all methods agree on the lower bound for the $68 \%$ confidence interval where $R(\Omega_m) \sim 0.6$, the FC method leads to noticeably smaller confidence intervals than either method based on graphical profile likelihoods. It should be noted that these methods require that $R(\Omega_m)$ adopts the same value at the extremities of the confidence intervals, but the FC method is free from the restriction\footnote{This restriction holds for $R$ in equation (\ref{eq:FC_ratio}) at the ends of the acceptance intervals in the confidence belt, but need not hold when the confidence intervals are plotted against the graphical profile likelihood.}. Moreover, when the FC confidence interval is mapped back to the graphical profile likelihood, it is clear that the area outside the confidence interval in the tails of the graphical profile likelihood is also not the same. It is interesting to also present the corresponding $R(H_0)$ for $z_{\textrm{min}}=0.5$ in Fig. \ref{fig:oh0_zmin05_profile} in order to document the kink in the curve at $H_0 \sim 40$ km/s/Mpc. This kink comes about due to the $\Omega_m \leq 1$ bound, which leads to a jump in $\chi^2 (\theta = H_0)$ in equation (\ref{eq:R}) and a corresponding smaller value of $R(H_0)$. Given that the kink happens outside of the Wilks' $68\%$ confidence interval, it has no bearing on it since Wilks' only requires a knowledge of the graphical profile likelihood in a local sense, but it affects the integration method that is globally sensitive. 

\begin{figure}[htb]
   \centering
\includegraphics[width=85mm]{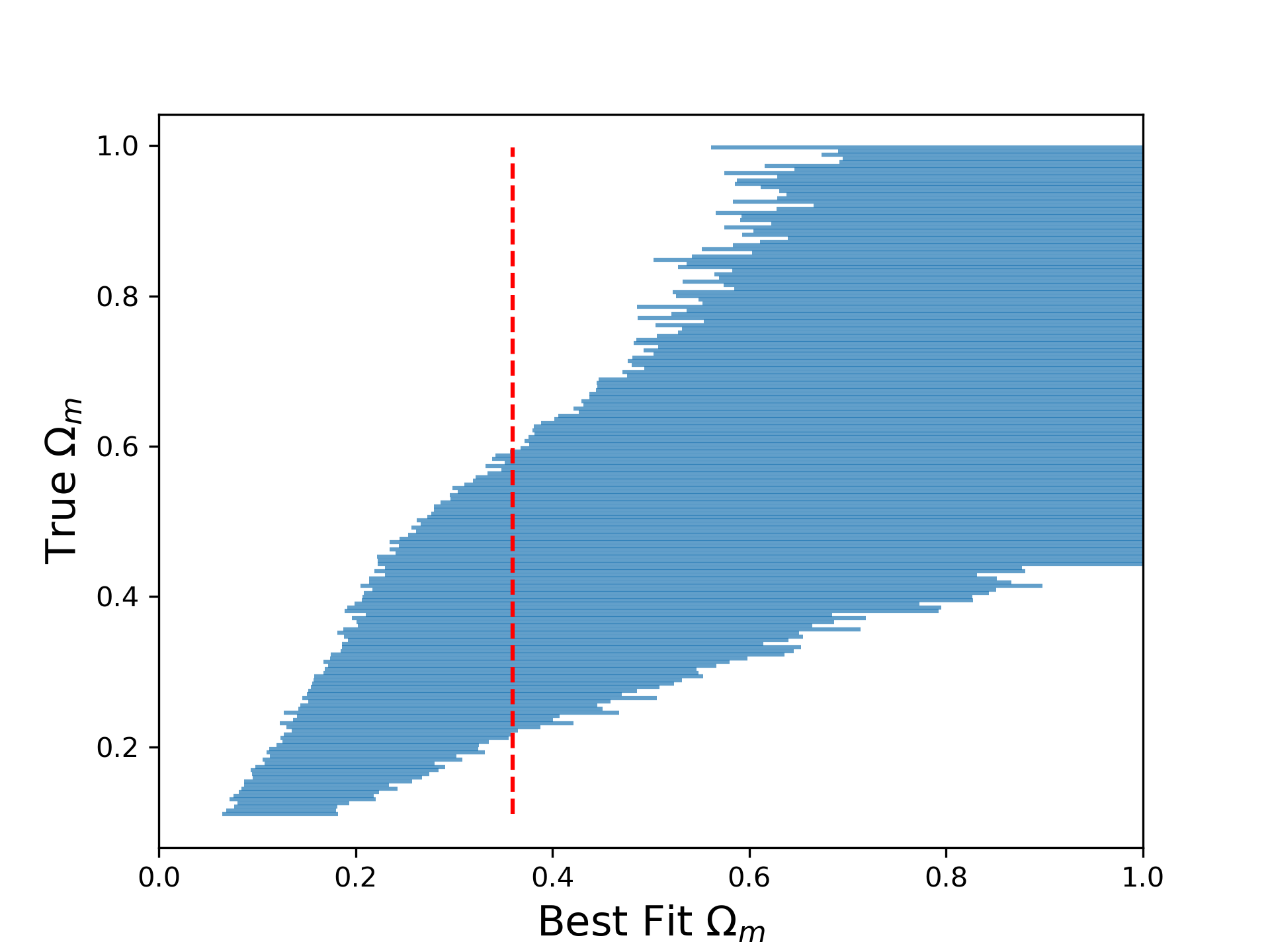} 
\caption{Same as Fig. \ref{fig:om_zmin00_belt} but for the CC data with $z_{\textrm{min}} = 0.5$.}
\label{fig:om_zmin05_belt}
\end{figure}

In Fig. \ref{fig:om_zmin05_belt} we also present the FC confidence belt for $\Omega_m$. In contrast to Fig. \ref{fig:h0_zmin00_belt} and Fig. \ref{fig:om_zmin00_belt} the end points of the acceptance intervals (horizontal lines) no longer trace two straight lines. Evidently, one is deep in a non-Gaussian regime. The confidence belts can be contrasted with Fig. 10 of \cite{Feldman:1997qc} and Fig. 11 of \cite{SPIDER:2021ncy}, where away from the boundary one recovers acceptance interval end points that trace parallel straight lines. In the top right corner of the plot there is visibly noise as the endpoints of the horizontal lines do not trace curves. This is expected as there are fewer MCMC configurations at larger $\Omega_m$, so one is performing fewer mocks. One could attempt to interpolate a curve to remove the noise, but since it is only pronounced far from the red dashed line, it does not affect results. Our confidence interval bears a closer resemblance to Fig. 6-8 of \cite{LiteBIRD:2023zmo} where the acceptance intervals get wider at larger values of the tensor-to-scalar ratio. Interestingly, despite the strong overlap of the papers on the tensor-to-scalar ratio \cite{SPIDER:2021ncy, LiteBIRD:2023zmo}, even in the setting where one fits a single parameter, the confidence intervals are qualitatively different.

\begin{figure}[htb]
   \centering
\includegraphics[width=85mm]{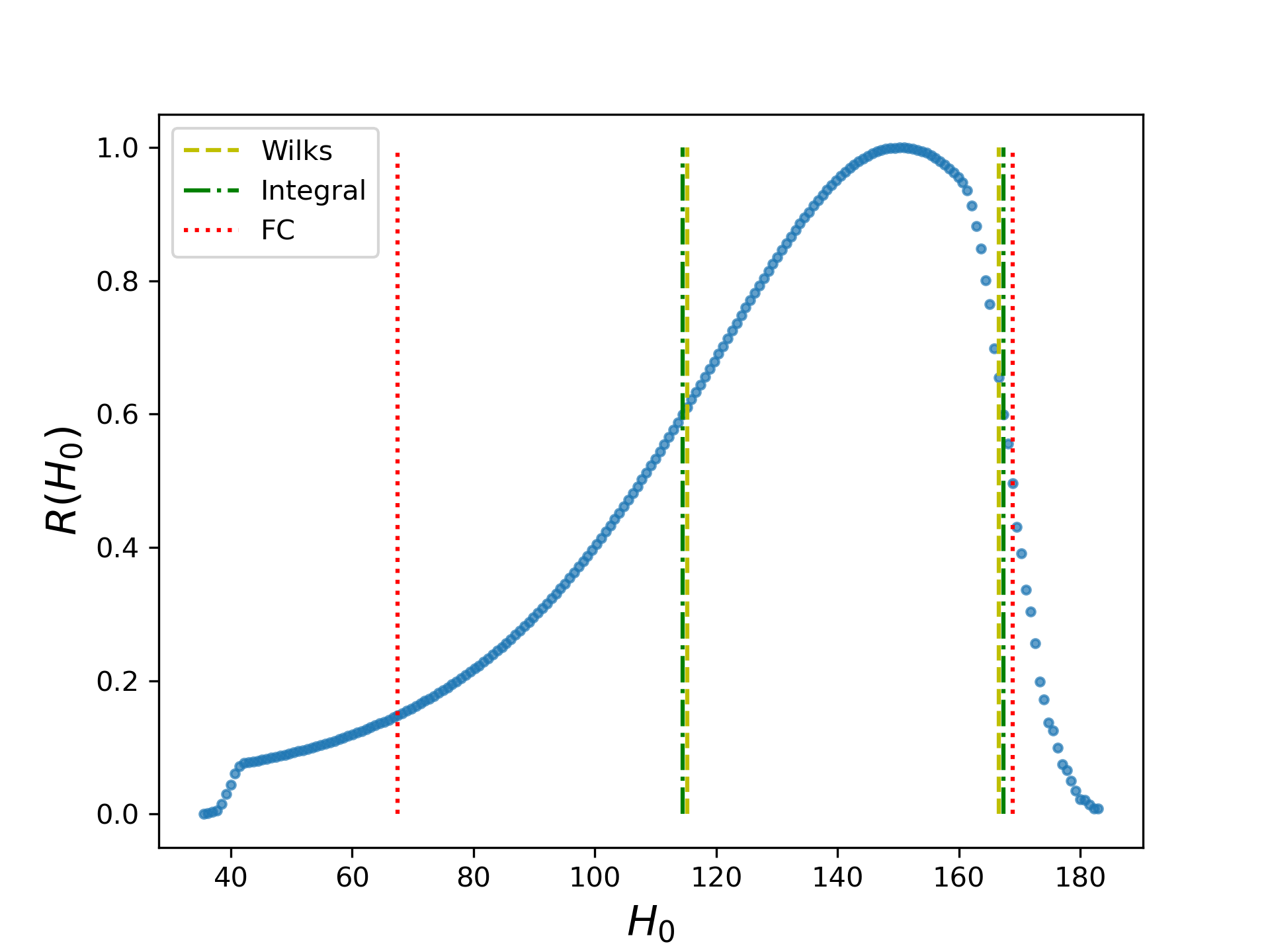} 
\caption{Same as Fig. \ref{fig:h0_zmin00_profile} but for the CC data with $z_{\textrm{min}} = 1$.}
\label{fig:h0_zmin10_profile}
\end{figure}

\begin{figure}[htb]
   \centering
\includegraphics[width=85mm]{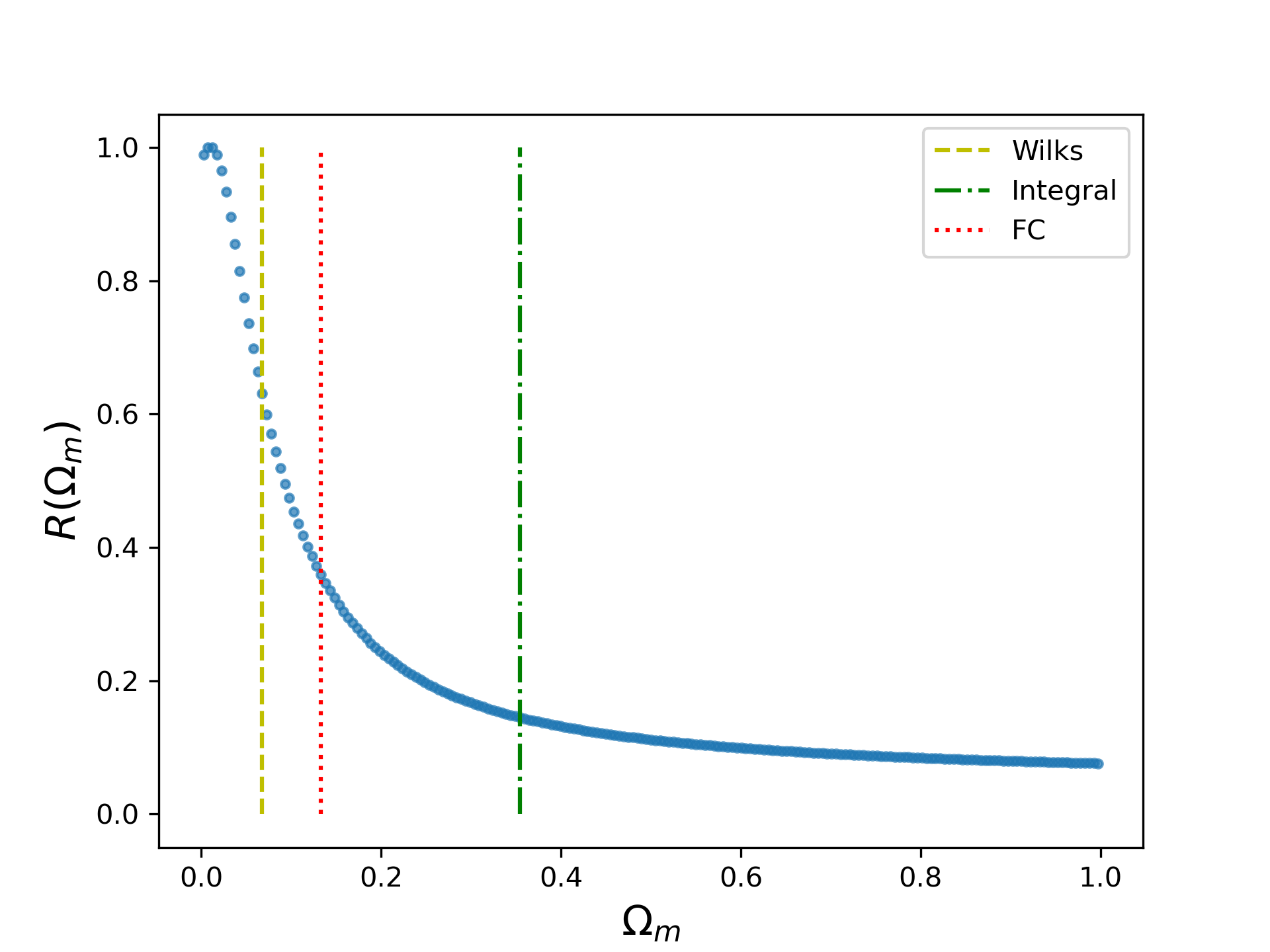} 
\caption{Same as Fig. \ref{fig:om_zmin00_profile} but for the CC data with $z_{\textrm{min}} = 1$.}
\label{fig:om_zmin10_profile}
\end{figure}

Given the similarities between the posteriors in Fig. \ref{fig:CC_posteriors} and \ref{fig:DESI_posteriors}, we do not expect to learn more from the mock DESI data apart from confirming that our results are not biased. This leaves us with one outlying situation to discuss, namely the red posterior in Fig. \ref{fig:CC_posteriors} corresponding to a zero physical matter density. In Fig. \ref{fig:h0_zmin10_profile} and Fig. \ref{fig:om_zmin10_profile} we plot the graphical profile likelihoods for $H_0$ and $\Omega_m$, respectively. Once again, the kink at lower $H_0$ values in $R(H_0)$ is due to the $\Omega_m \leq 1$ bound. 

In ref. \cite{Colgain:2023bge} based on i) graphical profile likelihoods and ii) $p$-values based on best fits to mock simulations (section 2.4 of \cite{Colgain:2023bge}), it was claimed that the $H_0 \sim 150$ km/s/Mpc, $\Omega_m \sim 0$ best fit values preferred by $z > 1$ CC data constituted a $\sim 2 \sigma $ shift away from the $H_0 \sim 70$ km/s/Mpc, $\Omega_m \sim 0.3$ values preferred by the full sample. We remind the reader that this result does not incorporate the CC covariance matrix \cite{Moresco:2020fbm}, so the shift is overestimated, but our interest is simply recovering an existing result in the literature. We further emphasise that the conclusion is supported by the two independent pillars i) and ii). In \cite{Colgain:2023bge}, it was argued that the $\sim 2 \sigma$ effect could be seen in the Integral confidence intervals in the $H_0$ graphical profile likelihood. However, upgrading the Integral (also Wilks) method to the FC method, we see that the canonical $H_0 \sim 70$ km/s/Mpc is within the $68 \%$ ($1 \sigma$) confidence interval. However, we still expect to see a $\sim 2 \sigma$ shift because of the mock simulations analysis. A hint of this is provided now by the $68 \%$ $\Omega_m$ confidence interval from the FC method in Fig. \ref{fig:om_zmin10_profile}, where we see that the confidence interval has contracted from the Integral to the FC method, thereby placing the $\Omega_m \sim 0.3$ value well outside of the $68 \%$ confidence interval \footnote{From Table \ref{tab:CC} we have $\Omega_m < 0.13$ for $z_{\textrm{min}} = 1$, where we observe that doubling this number gives us a ballpark limit for the $95 \%$ confidence interval upper bound.}. In the appendix we present the $95 \%$ confidence belt, which despite exhibiting noise is consistent with the $95 \%$ confidence interval terminating at $\Omega_m \sim 0.3$. As is clear from both the red $\Omega_m$ posterior in \ref{fig:CC_posteriors} and the $\Omega_m$ graphical profile likelihood in Fig. \ref{fig:om_zmin10_profile}, one is far from the Gaussian setting, so only the FC method can be fully trusted.

Our main results are summarised and quantified in Table \ref{tab:CC} and Table \ref{tab:DESI}, where we present a comparison of the $68 \%$ credible/confidence intervals for different redshift ranges across the observed CC and mock DESI datasets. In line with expectations, for the redshift ranges where the MCMC posteriors in Fig. \ref{fig:CC_posteriors} and Fig. \ref{fig:DESI_posteriors} are closest to Gaussian, we find excellent agreement across all methods. Noting that projection effects aside, only Bayesian and FC methods can be fully trusted as one ventures away from the large sample limit with Gaussian posteriors, one observes that the FC $68 \%$ confidence intervals are typically the same or smaller than Wilks/integration, thereby mirroring a result that one can prove in the large sample (Gaussian) setting, namely the FC method can only contract Wilks' confidence intervals when a boundary is present in the same limit \cite{Colgain:2024clf}. However, as is clear from Fig. \ref{fig:h0_zmin10_profile} and Fig. \ref{fig:om_zmin10_profile} it is easy to find situations where the converse is true and Wilks' theorem leads to smaller confidence intervals. One could worry that this is just an extreme example, but we have checked that DESI forecast data in the range $1.5 < z \leq 3.55$ (a combination of the last two bins in Table \ref{tab:DESI}) leads to the same result. Note, there is no contradiction with \cite{Colgain:2024clf} as the claimed result therein only applies to strictly Gaussian profile likelihoods in the presence of a boundary.

%\begin{figure}[htb]
%   \centering
%\includegraphics[width=85mm]{om_zmin_05_neyman_belt.png} 
%\caption{Same as Fig. \ref{fig:om_zmin00_belt} but for the CC data with $z_{\textrm{min}} = 0.5$.}
%\label{fig:om_zmin05_belt}
%\end{figure}

\begin{table}
    \centering
    \begin{tabular}{c|c|c|c|c}
    \rule{0pt}{3ex} $z_{\textrm{min}}$ & CC Count & Method & $H_0$ (km/s/Mpc) & $\Omega_m $\\
    \hline
    \rule{0pt}{3ex} \multirow{4}{*}{$0$}  &  \multirow{4}{*}{$33$} & MCMC & $67.8 \pm 3.1$ & $0.328^{+0.065}_{-0.056}$ \\
%    \cline{2-4}
    \rule{0pt}{3ex} & & Wilks & $68.2^{+3.0}_{-3.1}$ & $0.320^{+0.064}_{-0.053}$ \\
    \rule{0pt}{3ex} & & Integral & $68.2 \pm 3.1$ & $0.320^{+0.067}_{-0.056}$ \\
    \rule{0pt}{3ex} & & FC & $68.2^{+2.9}_{-3.3}$ & $0.320^{+0.064}_{-0.056}$ \\
    \hline 
    \rule{0pt}{3ex} \multirow{4}{*}{$0.25$}  & \multirow{4}{*}{$26$} & MCMC & $63.0^{+6.7}_{-7.2}$& $0.41^{+0.17}_{-0.11}$ \\
%    \cline{2-4}
    \rule{0pt}{3ex} & & Wilks & $65.1^{+6.3}_{-6.8}$ & $0.367^{+0.15}_{-0.097}$ \\
    \rule{0pt}{3ex} & & Integral & $65.1^{+6.5}_{-7.1}$ & $0.37^{+0.17}_{-0.11}$ \\
    \rule{0pt}{3ex} & & FC & $65.1^{+5.6}_{-9.0}$ & $0.367^{+0.17}_{-0.097}$ \\
    \hline
    \rule{0pt}{3ex} \multirow{4}{*}{$0.5$} & \multirow{4}{*}{$15$} & MCMC & $60^{+13}_{-11}$& $0.45^{+0.30}_{-0.19}$ \\
%    \cline{2-4}
    \rule{0pt}{3ex} & & Wilks & $66^{+13}_{-16}$ & $0.36^{+0.37}_{-0.15}$ \\
    \rule{0pt}{3ex} & & Integral & $66^{+12}_{-14}$ & $0.36^{+0.34}_{-0.15}$ \\
    \rule{0pt}{3ex} & & FC & $66^{+10}_{-14}$ & $0.36^{+0.24}_{-0.14}$ \\
     \hline
    \rule{0pt}{3ex} \multirow{4}{*}{$0.75$} & \multirow{4}{*}{$13$} & MCMC & $69^{+20}_{-17}$& $0.33^{+0.34}_{-0.18}$ \\
%    \cline{2-4}
    \rule{0pt}{3ex} & & Wilks & $81^{+18}_{-23}$ & $0.21^{+0.31}_{-0.10}$ \\
    \rule{0pt}{3ex} & & Integral & $81^{+17}_{-22}$ & $0.21^{+0.39}_{-0.11}$ \\
      \rule{0pt}{3ex} & & FC & $81^{+15}_{-25}$ & $0.21^{+0.24}_{-0.10}$ \\
     \hline
    \rule{0pt}{3ex} \multirow{4}{*}{$1$} & \multirow{4}{*}{$8$} & MCMC & $108^{+34}_{-44}$& $< 0.40$ \\
%    \cline{2-4}
    \rule{0pt}{3ex} & & Wilks & $150^{+16}_{-35}$ & $< 0.068$ \\
    \rule{0pt}{3ex} & & Integral & $150^{+17}_{-36}$ & $<0.36$ \\  
     \rule{0pt}{3ex} & & FC & $150^{+19}_{-83}$ & $< 0.13$ \\
    \end{tabular}
    \caption{A comparison of $68 \%$ Bayesian credible intervals and $68 \%$ frequentist confidence intervals across different methods as we restrict the CC dataset to redshifts $z > z_{\textrm{min}}$. We document the number of remaining data points through CC count.}
    \label{tab:CC}
\end{table}

Thus, we conclude that one cannot establish any hierarchy between FC confidence intervals and simpler Wilks' confidence intervals. Indeed, one cannot say that Wilks' method generically overestimates the FC method. Moreover, away from the Gaussian setting, the Integral method \cite{Gomez-Valent:2022hkb} need not agree with Wilks or FC confidence intervals. Nevertheless, it may still have value as a diagnostic of projection effects in MCMC posteriors.   

\begin{table}
    \centering
    \begin{tabular}{c|c|c|c}
    \rule{0pt}{3ex} $z$ & Method & $H_0$ (km/s/Mpc) & $\Omega_m $\\
    \hline
    \rule{0pt}{3ex} \multirow{4}{*}{$ 0.05 \leq z < 0.8$} & MCMC & $69.8^{+2.1}_{-2.2}$& $0.303^{+0.040}_{-0.036}$ \\
    \rule{0pt}{3ex} & Wilks & $70.0 \pm 2.1$ & $0.300^{+0.041}_{-0.036}$ \\
    \rule{0pt}{3ex} & Integral & $70.0^{+2.2}_{-2.1}$ & $0.300^{+0.041}_{-0.038}$ \\
    \rule{0pt}{3ex} & FC & $70.0^{+2.3}_{-1.9}$ & $0.300^{+0.041}_{-0.034}$ \\
    \hline 
    \rule{0pt}{3ex} \multirow{4}{*}{$0.8 < z < 1.5$} & MCMC & $69.1^{+4.4}_{-4.7}$& $0.311^{+0.067}_{-0.051}$ \\
%    \cline{2-4}
    \rule{0pt}{3ex} & Wilks & $70.0^{+4.3}_{-4.6}$ & $0.300^{+0.062}_{-0.048}$ \\
    \rule{0pt}{3ex} & Integral & $70.0^{+4.5}_{-4.6}$ & $0.300^{+0.068}_{-0.051}$ \\
    \rule{0pt}{3ex} & FC & $70.0^{+3.9}_{-5.2}$ & $0.300^{+0.065}_{-0.045}$ \\
    \hline
    \rule{0pt}{3ex} \multirow{4}{*}{$1.5 < z < 2.3$} & MCMC & $60^{+18}_{-14}$& $0.43^{+0.33}_{-0.19}$ \\
%    \cline{2-4}
    \rule{0pt}{3ex} & Wilks & $70^{+18}_{-24}$ & $0.30^{+0.46}_{-0.13}$ \\
    \rule{0pt}{3ex} & Integral & $70^{+15}_{-20}$ & $0.30^{+0.39}_{-0.12}$ \\
    \rule{0pt}{3ex} & FC & $70^{+14}_{-19}$ & $0.30^{+0.21}_{-0.12}$ \\
     \hline
    \rule{0pt}{3ex} \multirow{4}{*}{$2.3 < z \leq 3.55$} & MCMC & $59^{+28}_{-15}$& $0.43^{+0.35}_{-0.25}$ \\
    \rule{0pt}{3ex} & Wilks & $70^{+45}_{-31}$ & $>0.098$ \\
    \rule{0pt}{3ex} & Integral & $70^{+27}_{-31}$ & $0.30^{+0.49}_{-0.12}$ \\
    \rule{0pt}{3ex} & FC & $70^{+27}_{-31}$ & $0.30^{+0.34}_{-0.19}$ \\
    \end{tabular}
    \caption{A comparison of $68 \%$ Bayesian credible intervals and $68 \%$ frequentist confidence intervals across different methods as we restrict the DESI forecast dataset to four redshift bins with approximately the same number of data points (7 or 8) in each bin.}
    \label{tab:DESI}
\end{table}

We finish with comments on the extent to which MCMC credible intervals and Wilks' confidence intervals overlap with FC confidence intervals. We limit the comparison between MCMC credible intervals and FC confidence intervals to the lowest redshift bins as in high redshift bins the 1D MCMC posteriors are heavily impacted by projection effects, and having diagnosed the projection effects, it is no longer a fair comparison. From the first two rows in Table \ref{tab:CC} and Table \ref{tab:DESI}, we find that MCMC credible intervals correspond to $95 \% - 105 \%$ and $100 \% - 107\%$ of the FC confidence intervals, respectively. Good agreement is expected and we confirm that this is the case. Performing the same comparison between the Wilks and FC methods at low redshifts confirms that the Wilks' confidence intervals correspond to $90 \%-98 \%$ and $98 \%-103 \%$ of the FC confidence intervals in Table \ref{tab:CC} and Table \ref{tab:DESI}, respectively. Overall, the agreement is no worse than $10 \%$ for posteriors and profile likelihoods that are close to Gaussian. One can attempt to extend the comparison between Wilks and FC to all the redshift ranges, where it should be borne in mind that Wilks is at best an approximation applied beyond its range of validity. Doing so, we find that the Wilks' confidence intervals correspond to $50 \% - 137 \%$ and $127 \% -179\%$ of the FC confidence intervals in Table \ref{tab:CC} and Table \ref{tab:DESI}. Evidently, there is a danger that the Wilks method may underestimate or overestimate the FC method by a factor close to two, representing a serious discrepancy, especially if translated into a $\Lambda$CDM tension.

\section{Discussion}
The motivation of this paper is to study both graphical profile likelihoods (Wilks' theorem \cite{Wilks}, integration \cite{Gomez-Valent:2022hkb}) and more generic profile likelihood methods based on simulations \cite{Feldman:1997qc} in a setting where Bayesian MCMC posteriors are non-Gaussian and subject to projection effects. We do this for observational $H(z)$ constraints from CC data \cite{Moresco:2023zys} and forecast mock DESI data \cite{DESI:2016fyo} fitted to the flat $\Lambda$CDM model, where as noted in \cite{Colgain:2022tql}, the removal of anchoring low redshift data leads to non-Gaussian distributions as the $\Lambda$CDM model transitions from a 2-parameter model at lower redshifts to an effective 1-parameter model at higher redshifts. This feature of the $\Lambda$CDM model applies to $H(z)$ and $D_{M}(z)$ constraints whenever not combined, \textit{cf.} anisotropic BAO.  

The big picture here is that it is imperative to perform consistency checks of the $\Lambda$CDM model in response to $\Lambda$CDM tensions in order to ascertain whether the fitting parameters of the model are constant or not \cite{Akarsu:2024qiq}. This necessitates tomographic analyses where one bins the data by redshift, which invariably leads to smaller samples, potentially non-Gaussian distributions and subtleties in identifying Bayesian credible or frequentist confidence intervals. In particular, as observed in ref. \cite{Colgain:2023bge}, observed high redshift $z > 1$ CC data \cite{Moresco:2023zys} is more consistent with a constant $H(z)$ ($\Omega_m = 0$ in $\Lambda$CDM) than a $H(z)$ increasing with redshift ($\Omega_m > 0$ in $\Lambda$CDM). One of the goals of this paper is to generalise the graphical methods in \cite{Colgain:2023bge}, which may only hold for Gaussian distributions, to the proper non-Gaussian setting. 

The appropriate way to do this is to the follow the FC prescription \cite{Feldman:1997qc}, a generalisation of the Neyman belt construction \cite{Neyman:1937uhy} that incorporates parameters with boundaries. Even with the FC prescription, subtleties arise. Interesting situations in cosmology typically involve models with multi-dimensional parameter spaces. Typically one assumes the same underlying cosmology with fixed input parameters when one mocks up the simulations \cite{SPIDER:2021ncy, LiteBIRD:2023zmo}. Moreover, as noted in \cite{LiteBIRD:2023zmo}, the confidence intervals themselves are sensitive to the number of parameters one fits back to the mocks, thus pointing to an inherent arbitrariness. As explained in the text, fixing parameters impacts both the best fit values in the acceptance intervals and the likelihood ratios that dictate whether a best fit is included in the acceptance interval or not. Despite the restricted setting in this study with 2-parameters, our approach is completely general. We recycle the $(H_0, \Omega_m)$ values in the MCMC chain as the basis for the mocks and fit back both parameters when executing the FC prescription. When fitting back $\Omega_m$ we still impose $\Omega_m \in [0,1]$, but have checked that relaxing the upper bound to allow $\Omega_m > 0$ best fits, a mathematically allowed process (allowing $\Omega_m < 0$ is not), has little bearing on our results. 

Our main observation is that all methods show good agreement in the close to Gaussian regime, which is the expected result. Nevertheless, beyond the Gaussian regime, one needs to employ the Neyman belt \cite{Neyman:1937uhy} or FC prescription \cite{Feldman:1997qc}, where we find that approximation with simpler graphical profile likelihoods may be poorer. Concretely, we find that the confidence intervals may be overestimated or underestimated by a factor of 2. We also note that in contrast to the strict Gaussian limit \cite{Colgain:2024clf}, there is no clear hierarchy in the size of confidence intervals across frequentist methods. Finally, we observe that the $\sim 2 \sigma$ shift in cosmological parameters across the CC dataset noted in ref. \cite{Colgain:2023bge} appears to be driven by a discrepancy in the $\Omega_m$ parameter and not the $H_0$ parameter as one naively concludes when approximating profile likelihoods with purely graphical methods that only hold in the Gaussian regime. 

As we show in the appendix, fixing one of the parameters would radically change this conclusion by restoring close to Gaussian behaviour in an inherently non-Gaussian setting and greatly shrinking the confidence intervals. When taken in tandem with the observation from ref. \cite{LiteBIRD:2023zmo} that the confidence intervals depend on the number of fitted parameters, fixing parameters runs the risk of underestimated frequentist confidence intervals. We conclude that it is safest to mock all parameters and fit back all parameters when constructing the FC confidence intervals.

\section*{Acknowledgements}
We thank Paolo Campeti, Robert Cousins and Laura Herold for helpful correspondence on profile likelihood methods. This article/publication is based upon work from COST Action CA21136 – “Addressing observational tensions in cosmology with systematics and fundamental physics (CosmoVerse)”, supported by COST (European Cooperation in Science and Technology). SB would also like to extend his gratitude to the University Grants Commission (UGC), Govt. of India for their continuous support through the Junior Research Fellowship.

\appendix

\section{Confirming MCMC degeneracy}
In this section we confirm that the high redshift degeneracy (banana-shaped contour) evident in Fig. \ref{fig:CC_posteriors} (also Fig. \ref{fig:DESI_posteriors}) cannot be removed by changing the MCMC algorithm. Note, this is the expected result, since it is intuitive for MCMC practitioners that exclusively high redshift data in the late Universe struggles to break the degeneracy between $H_0$ and $\Omega_m$ in the flat $\Lambda$CDM model. We focus on the CC data, but increase the low redshift cutoff to $z_{\textrm{min}} = 1.2$, which is sufficient to remove the potentially confusing secondary mode from the bimodal $z_{\textrm{min}}=1$ $H_0$ posterior in Fig. \ref{fig:CC_posteriors}. 

\begin{figure}[htb]
   \centering
\includegraphics[width=80mm]{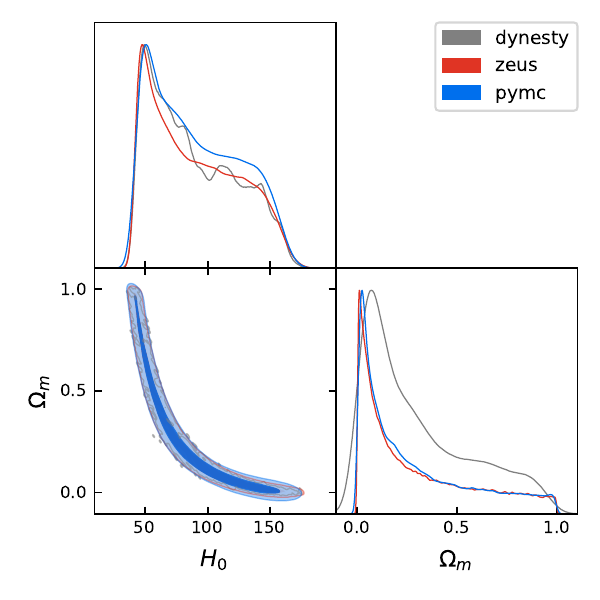} 
\caption{Confirmation that changing the Bayesian algorithm for CC data with $z_{\textrm{min}} = 1.2$ does not change Bayesian posteriors appreciably.}
\label{fig:zmin12_degeneracy}
\end{figure}

In Fig. \ref{fig:zmin12_degeneracy} we present results from \textit{dynesty} \cite{Speagle:2019ivv}, \textit{zeus} \cite{Karamanis:2021tsx} and \textit{pymc} \cite{Abril-Pla_PyMC_a_modern_2023}. What the reader should take away is that the 2D posterior follows a constant $H_0 \sqrt{\Omega_m}$ curve in the $(H_0, \Omega_m)$-plane and the posterior is curtailed by the $\Omega_m \in [0, 1]$ priors, thus the result is prior dependent. Relaxing the $\Omega_m \leq 1$ bound allows the posterior to stretch further, leading to a peak in the projected $H_0$ posterior at lower values (see for example Fig. 2 of \cite{Malekjani:2023ple}). There is also a noticeable bump in the $H_0$ posterior at $H_0 \sim 150$ km/s/Mpc, which is all that remains of the best fit $H_0$ value \cite{Colgain:2023bge}. Given the pronounced degeneracy and projection effect, the only safe conclusion a Bayesian cosmologist should draw is that the data fails to separate $H_0$ and $\Omega_m$, but only constrains the combination $H_0 \sqrt{\Omega_m}$; its value can be extracted from the curve in the $(H_0, \Omega_m)$-plane.

\section{Comment on $95 \%$ Confidence Interval}
\begin{figure}[htb]
   \centering
\includegraphics[width=80mm]{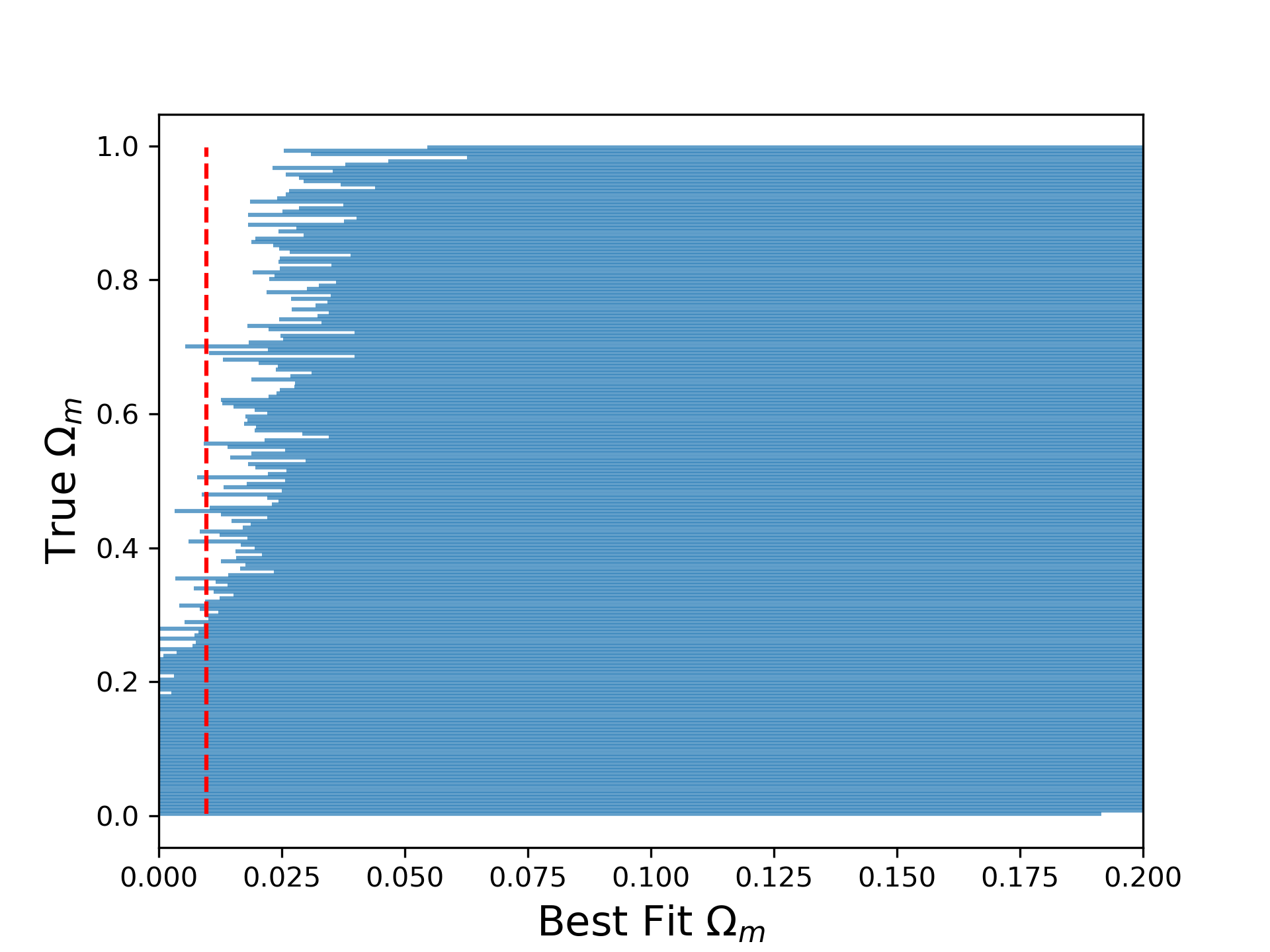} 
\caption{The $95 \%$ confidence belt for the $\Omega_m$ parameter and CC data with $z_{\textrm{min}} = 1$.}
\label{fig:95_interval}
\end{figure}

All of our analysis in the paper focuses on $68 \%$ confidence intervals. As explained in the text, in \cite{Colgain:2023bge} an approximate $2 \sigma$ or $95 \%$ confidence level shift in the cosmological parameters was claimed in the CC dataset in Fig. \ref{fig:CCdata} between the full sample and a high redshift subsample with $z > 1$. From analysis in the text, one sees that this discrepancy is only evident in the $\Omega_m$ parameter. See Fig. \ref{fig:om_zmin10_profile} where only $68 \%$ confidence intervals are presented. In Fig. \ref{fig:95_interval} we show the confidence belt one gets when one employs the FC method with $\theta = \Omega_m$ and retains $\theta_{\textrm{MLE}}$ values in each bin corresponding to $95 \%$ of the mock simulations with the largest $R$ from equation (\ref{eq:FC_ratio}).   
 
From Fig. \ref{fig:95_interval} it is clear that there is considerable noise in the confidence belt. The reader can compare to Fig. \ref{fig:om_zmin05_belt} where one only encounters similar noise in the top right corner of the plot. The real obstacle here is not the MCMC chain, it is the time it takes to construct the mocks and fit the parameters. One can in principle reduce the noise by running a longer MCMC chain. The main observation to take away from Fig. \ref{fig:95_interval} is that despite the noise beyond the true value of $\Omega_m \sim 0.3$, the red dashed line has exited from the majority of the acceptance intervals (horizontal lines) and is only dragged back into the acceptance intervals by isolated acceptance intervals. This provides support for the observation in \cite{Colgain:2023bge} that this is a shift close to the $2 \sigma$ or $95 \%$ confidence level.  

\section{One Parameter Fit}
In the body of this work we have done the most general thing. We allowed both parameters $(H_0, \Omega_m)$ to vary in the injected cosmology and we fitted back both parameters. In the literature \cite{SPIDER:2021ncy, LiteBIRD:2023zmo} it is common to use the same injected cosmology and make choices about the number of parameters one fits back. As observed in \cite{LiteBIRD:2023zmo} the confidence intervals one finds may depend on the number of fitting parameters one fits back. We will demonstrate how this affects our results by fixing $\Omega_m$. Concretely, we consider the CC data with $z_{\textrm{min}} = 1$ with best fit parameters $H_0 = 150.4$ km/s/Mpc and $\Omega_m = 0.0098$. As is clear from the posterior in Fig. \ref{fig:CC_posteriors}, or the graphical profile likelihoods in Fig. \ref{fig:h0_zmin10_profile} and Fig. \ref{fig:om_zmin10_profile}, one is deep in the non-Gaussian regime. 

\begin{figure}[htb]
   \centering
\includegraphics[width=80mm]{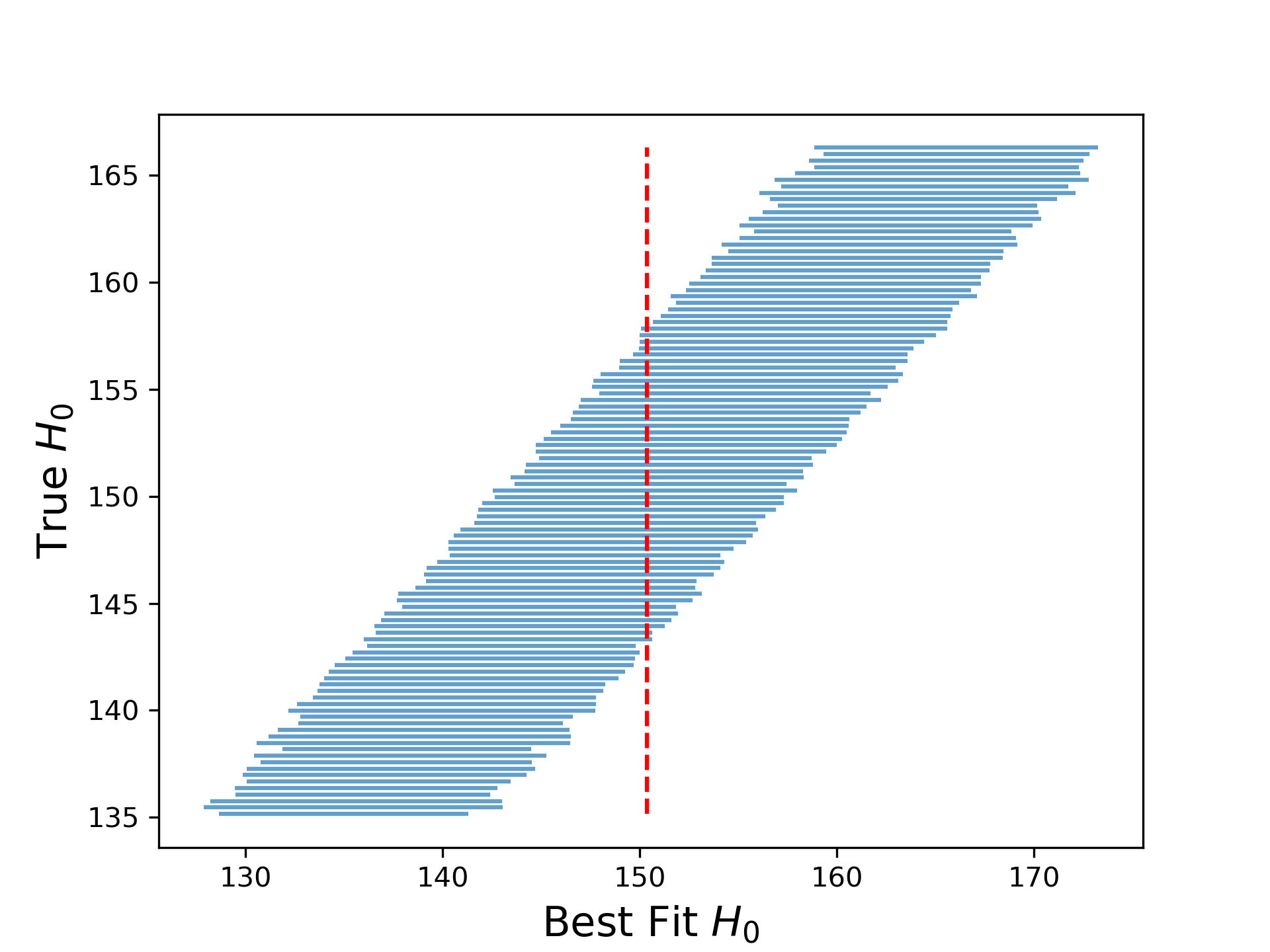} 
\caption{The $68 \%$ confidence belt for the $H_0$ parameter and CC data with $z_{\textrm{min}} = 1$ and $\Omega_m$ fixed to its best fit value throughout.}
\label{fig:1param}
\end{figure}

We fix $\Omega_m = 0.0098$ and fit $H_0$ to the data with MCMC leading to $H_0 = 150.3 \pm 7.4$ km/s/Mpc. It is worth noting that the errors are small relative to the 2-parameter fits in Table \ref{tab:DESI} and are symmetric, so whatever non-Gaussianity was present has been lost. We now repeat the steps in the main text by binning the $H_0$ values in the MCMC chain, constructing mock simulations and fitting back only $H_0$ with fixed $\Omega_m$. Note, $\Omega_m$ is fixed in the mocks and the fits. In Fig. \ref{fig:1param} we present the $68 \%$ $H_0$ confidence interval, which gives us the result $H_0 = 150.4^{+7.5}_{-7.0}$ km/s/Mpc. On the assumption of Gaussian confidence intervals, this places $H_0 = 70$ km/s/Mpc at $ \sim 8 \sigma$, thus greatly overestimating the $\sim 2 \sigma$ significance seen in $p$-values based on mock simulations \cite{Colgain:2023bge}. This can be contrasted with the result from the 2-parameter fit $H_0 = 150^{+19}_{-83}$ km/s/Mpc from Table \ref{tab:CC}. The problem here is that when we fix $\Omega_m$ we lose all trace of inherent non-Gaussianity that is present. 

As an added check, we can repeat the process for the full CC sample with $z_{\textrm{min}} = 0$, which is close to the Gaussian regime. Fixing $\Omega_m$ to its best fit value, we find $H_0 = 68.1^{+1.6}_{-1.5}$ km/s/Mpc. Comparing to Table \ref{tab:CC} and $H_0 = 67.8 \pm 3.1$ km/s/Mpc, we see that fixing $\Omega_m$ decreases the confidence intervals by a factor of 2. From the FC method, we find $H_0 = 68.2^{+1.5}_{-1.6}$ km/s/Mpc, thus agreeing well with MCMC under the same assumptions of fixed $\Omega_m$. What we learn here is that irrespective of whether one is in a Gaussian regime or not, fixing parameters leads to smaller confidence intervals.

\bibliography{refs}

\end{document}